\documentclass{aa}  

\usepackage{graphicx}
\usepackage{txfonts}
\usepackage{pdflscape}
\usepackage{comment}
\usepackage{orcidlink}
\usepackage[switch]{lineno}
\usepackage{hyperref}
\usepackage{multirow}

\newcommand{\Hydrogen}{\ion{H}{}}
\newcommand{\Helium}{\ion{He}{}}
\newcommand{\HeliumI}{\ion{He}{I}}
\newcommand{\Halpha}{\ion{H}{$\alpha$}}
\newcommand{\Hbeta}{\ion{H}{$\beta$}}

\newcommand{\HeII}{\ion{He}{II}}

\newcommand{\SIIF}{[\ion{S}{II}]}
\newcommand{\NII}{\ion{N}{II}}
\newcommand{\NI}{\ion{N}{I}}
\newcommand{\KI}{\ion{K}{I}}
\newcommand{\CaII}{\ion{Ca}{II}}
\newcommand{\Calcium}{\ion{Ca}{}}
\newcommand{\MgI}{\ion{Mg}{I}}
\newcommand{\NaID}{\ion{Na}{ID}}
\newcommand{\ztfg}{$g$}
\newcommand{\ztfr}{$r$}
\newcommand{\ztfi}{$i$}

\begin{document} 

   \title{A real-time search for Type Ia Supernovae with late-time CSM interaction in ZTF}

   \author{Jacco H. Terwel \orcidlink{0000-0001-9834-3439} \inst{1,2}
   \and Kate Maguire \orcidlink{0000-0002-9770-3508} \inst{1} 
   \and Se{\'a}n J. Brennan \orcidlink{0000-0003-1325-6235} \inst{3}
   \and Lluís Galbany \orcidlink{0000-0002-1296-6887} \inst{4,5}
   \and Simeon Reusch \orcidlink{0000-0002-7788-628X} \inst{6}
   \and Steve Schulze \orcidlink{0000-0001-6797-1889} \inst{7}
   \and Niilo Koivisto \orcidlink{0009-0007-7151-7313} \inst{2,8}
   \and Tapio Pursimo \orcidlink{0000-0002-5578-9219} \inst{2,9}
   \and Samuel Grund S\o rensen \orcidlink{0009-0009-5887-4281} \inst{2,10}
   \and María Alejandra Díaz Teodori \orcidlink{0009-0002-1852-7671} \inst{2,8}
   \and Astrid Guldberg Theil \inst{2,10}
   \and Mikael Turkki \orcidlink{0009-0009-1581-1408} \inst{2,11,8,12}
   \and Tomás E. Müller-Bravo \orcidlink{0000-0003-3939-7167} \inst{1,13}
   \and Umut Burgaz \orcidlink{0000-0003-0126-3999} \inst{1}
   \and Young-Lo Kim \orcidlink{0000-0002-1031-0796} \inst{14}
   \and Joshua S. Bloom \orcidlink{0000-0002-7777-216X} \inst{15,16}
   \and Matthew J. Graham \orcidlink{0000-0002-3168-0139} \inst{17}
   \and Mansi M. Kasliwal \orcidlink{0000-0002-5619-4938} \inst{17}
   \and Shri R. Kulkarni \orcidlink{0000-0001-5390-8563} \inst{17}
   \and Frank J. Masci \orcidlink{0000-0002-8532-9395} \inst{18}
   \and Josiah Purdum \orcidlink{0000-0003-1227-3738} \inst{19}
   \and Oleksandra Pyshna \inst{17}
   \and Avery Wold \orcidlink{0000-0002-9998-6732} \inst{18}
    }

   \institute{School of Physics, Trinity College Dublin, The University of Dublin, Dublin 2, Ireland\\
   \email{terwelj@tcd.ie}
   \and Nordic Optical Telescope, Rambla José Ana Fernández Pérez 7, ES-38711 Breña Baja, Spain
   \and The Oskar Klein Centre, Department of Astronomy, Stockholm University, AlbaNova, SE-10691 Stockholm, Sweden
   \and Institute of Space Sciences (ICE-CSIC), Campus UAB, Carrer de Can Magrans, s/n, E-08193 Barcelona, Spain.
   \and Institut d'Estudis Espacials de Catalunya (IEEC), 08860 Castelldefels (Barcelona), Spain
   \and Leibniz Institute for Astrophysics, An der Sternwarte 16, 14482 Potsdam, Germany
   \and Center for Interdisciplinary Exploration and Research in Astrophysics (CIERA), Northwestern University, 1800 Sherman Ave., Evanston, IL 60201, USA
   \and Department of Physics and Astronomy, University of Turku, Vesilinnantie 5, Turku FI-20014, Finland
   \and Department of Physics and Astronomy, Aarhus University, Munkegade 120, 8000, Aarhus C, Denmark
   \and Alum of Department of Physics and Astronomy, Aarhus University, Ny Munkegade 120, 8000 Aarhus C, Denmark
   \and Finnish Centre for Astronomy with ESO (FINCA), University of Turku, FI-20014 Turku, Finland
   \and Aalto University Metsähovi Radio Observatory, Metsähovintie 114, 02540 Kylmälä, Finland
   \and Instituto de Ciencias Exactas y Naturales (ICEN), Universidad Arturo Prat, Chile
   \and Department of Astronomy \& Center for Galaxy Evolution Research, Yonsei University, Seoul 03722, Republic of Korea
   \and Department of Astronomy, University of California, Berkeley, 501 Campbell Hall, Berkeley, CA 94720, USA
   \and Lawrence Berkeley National Laboratory, 1 Cyclotron Road MS 50B-4206, Berkeley, CA, 94720, USA
   \and Division of Physics, Mathematics, and Astronomy, California Institute of Technology, Pasadena, CA 91125, USA
   \and IPAC, California Institute of Technology, 1200 E. California Blvd, Pasadena, CA 91125, USA
   \and Caltech Optical Observatories, California Institute of Technology, Pasadena, CA 91125
   }

   \date{Received XXX; accepted YYY}
 
  \abstract 
   {While it is generally accepted that Type Ia supernovae (SNe Ia) are the terminal explosions of white dwarves (WDs), the nature of their progenitor systems and the mechanisms that lead up to these explosions are still widely debated. In rare cases the SN ejecta interact with circumstellar material (CSM) that was ejected from the progenitor system some time before the SN. The longer the delay between the CSM creation and the explosion, the larger the distance between the explosion site and the CSM, and the later the interaction starts after the SN. The unknown distance between the CSM and SN explosion site makes it impossible to predict when the interaction will start. If the time between the SN and start of CSM interaction is of the order of months to years the SN has generally faded and is not actively followed up anymore, making it even more difficult to detect the interaction while it happens. Here we report on a real-time monitoring program which ran between 13 November 2023 and 9 July 2024, monitoring 6\,914 SNe Ia for signs of late-time rebrightening using the Zwicky Transient Facility (ZTF). Flagged candidates were rapidly followed up with photometry and spectroscopy to confirm the late-time excess and its position. We report the discovery of a $\sim50$ day rebrightening event in SN~2020qxz around 1\,200 rest frame days after the peak of its light curve. SN~2020qxz had signs of early CSM interaction but had faded from view over two years before its reappearance. Follow-up spectroscopy revealed the presence of four emission lines that faded shortly after the end of the ZTF detected rebrightening event. Our best match for these emission lines are \Hbeta\ (blue shifted by $\sim5\,900$~km~s$^{-1}$) and \CaII$_{\lambda8\,542}$, \NI$_{\lambda8\,567}$, and \KI$_{\lambda\lambda 8\,763, 8\,767}$, all blue shifted by 5\,100~km~s$^{-1}$ (although we note that the line identifications are uncertain). This shows that catching and following up on late-time interactions as they occur can give new clues about the nature of the progenitor systems that produce these SNe by putting constraints on the possible type of donor star, and the only way to do this systematically is to use large sky surveys such as ZTF and the upcoming Vera C. Rubin Observatory’s Legacy Survey of Space and Time (LSST) to monitor a large sample of objects for the rare events that reappear long after it faded from view.}


   \maketitle

\section{Introduction}
Type Ia supernovae (SNe Ia) are the terminal explosions of white dwarfs (WDs) and are well-known as standardisable candles that allow us to estimate their distance based on the light curve properties around their peak after scaling them using simple relations \citep[e.g.][]{Phillips_rel, colour_corr, Phillips_rel2}. Despite decades of study the exact nature the progenitor systems and mechanisms that lead up to these explosions remain unknown. A fraction of SNe Ia have been found to be photometrically and/or spectroscopically different, and many subclasses have been found that are not standardisable like the normal SNe Ia \citep[see][for recent reviews]{Taubenberger_HbSNe, DR2_diversity}. These differences hold clues and can give insights into the progenitor systems before they exploded.

One such subclass are SNe Ia which are interacting with circumstellar material (CSM), commonly known as SNe Ia-CSM or SNe Ia-02ic after the first discovered member of this group \citep[SN 2002ic,][]{02ic_H_det, Hamuy_02ic}. This interaction is often characterised by the presence of narrow Balmer emission lines in the spectra. As hydrogen is typically not present in Type I SNe, these lines are interpreted as interaction between the SN ejecta and \Hydrogen-rich material that was ejected from the progenitor system some time before the explosion. One notable exception to this is SN 2020eyj, which \citet{Kool_He_CSM} present as a Ia-CSM interacting with \Helium-rich material coming from a \Helium star and a WD binary progenitor system.

SN 2011km (PTF11kx) is a well-observed SN~Ia-CSM which showed signs of a complex CSM consisting of multiple shells \citep{ptf11kx}. At early times narrow absorption lines revealed the presence of CSM near the SN, but the emergence of \Hydrogen and \Calcium emission lines marked the onset of CSM interaction. As these emission lines were at a different velocity compared to the absorption features showed that the interacting CSM was a different shell than the absorbing material. \citet{ptf11kx} suggested that the presence of multiple shells points to a symbiotic nova progenitor system.

CSM interaction lines only emerge after the ejecta reach the CSM, resulting in a delay that depends on the distance between the explosion and the material. Usually this is within the first few weeks to months after the explosion. The interaction acts as a new light source and can significantly alter the broadband evolution of the light curve \citep[see e.g.][]{Ia-CSM_BTS}. In some cases this is the first observed sign of the interaction as the SN might previously already have been classified as e.g. a normal SN~Ia. For instance, SN 2023ggb was classified as a SN~Ia on 23-04-2023 \citep{2023ggb_classif1} but reclassified as a SN~Ia-CSM on 03-06-2023 \citep{2023ggb_classif2} after its unusual light curve evolution warranted extra follow-up observations.

This poses a challenge for studying SNe~Ia-CSM. It is a rare subclass with only $\sim35$ members \citep{2005gj, Ia-CSM_Silverman, Ia-CSM_BTS} and its characteristic features only emerge at a later stage, possibly when it is no longer actively being followed up on. Moreover, a rebrightening event due to late-time CSM interaction could occur months or even years after the SN has completely faded from view. The only way to systematically search for such events is by continuously following a large sample of known older objects.

\citet{2015cp} performed a systematic targeted search for late-time CSM interaction in $\geq1$ year old SNe Ia using the \textit{Hubble Space Telescope} (HST), focusing on subclasses that are associated with CSM interaction such as SNe~Ia-91T \citep{Ia-CSM_and_91T_connection}. They discovered late-time CSM interaction in SN 2015cp at 664 days after the peak (although the last observation without signs of CSM interaction was at 45 days after the peak). This showed the existence of objects where CSM interaction starts at late times. \citet{GALEX_Late_CSM} looked for late-time CSM interaction in a sample of 1\,080 SNe~Ia using archival UV-band data from the \textit{Galactic Evolution Explorer} (GALEX). While four SNe were detected in the UV near their peak, none showed signs of late-time CSM interaction. Both studies conclude that late-time CSM interaction with the strength of that seen in SN 2015cp is rare, occurring in $<5\%$ of SNe~Ia.

Large sky surveys such as the Asteroid Terrestrial-impact Last Alert System \citep[ATLAS,][]{ATLAS} and Zwicky Transient Facility \citep[ZTF,][]{ZTF_Surveys_Scheduler, ZTF_overview_and_1st_results, ZTF_Science_Objectives, ZTF_Instrumentation, ZTF_Observing_System} observe large parts of the night sky on a regular basis for long periods of time. This is ideal for systematically discovering new transients that are then automatically followed up photometrically. Through this it is possible to build a large catalogue of all transients that occurred in a certain part of the sky over a certain time period, down to a certain limiting magnitude. ZTF covers the entire northern hemisphere above declination~$\sim -30^{\circ}$ with a total covered area between 25\,000 to 30\,000 square degrees, observing in three broadband optical filter (ZTF-\ztfg, ZTF-\ztfr, and ZTF-\ztfi) with a two to three night cadence and a $\sim 20.5$ mag limit per single exposure since March 2018. Every point covered by ZTF has therefore a light curve that spans years and includes observations from before, during, and after a transient was visible (assuming the transient occurred after the start of ZTF).

In \citet{JHT_DR2}, the ZTF SN Ia DR2 sample \citep[][Smith et al.~in prep.]{DR2_Overview}, which covers all ZTF observed SNe~Ia that were first detected before 2021, was examined for faint signals occurring $>100$ days after the peak of the SN. To do this systematically, they developed a binning pipeline that averages observations together and showed that with this technique it is possible to detect signals up to $\sim 1$ mag beyond the single exposure magnitude limit by sacrificing time sensitivity. Using this, \citet{JHT_DR2} found three candidates out of a sample of 3628 SNe Ia with late-time signals that could be explained by late-time CSM interaction. By simulating the survey they showed that this corresponds to an absolute rate of $8_{-4}^{+20}$ to $54_{-26}^{+91}$ Gpc$^{-3}$ yr$^{-1}$ assuming a constant SN Ia rate for $z \leq 0.1$ \citep{SNIa_rate}.

In \citet{JHT_pre-ZTF}, this pipeline was adapted to search for late-time signals in any transient discovered between 2008 and 2018 whose position is covered by ZTF. The ZTF light curves of these objects cover up to 10 to 15 years after the explosion, which allowed them to search for signals at even later epochs. Their sample included 4\,991 SNe~Ia, and 2\,727 other transients. While 98 objects had detections in the binned ZTF data, most of these were sibling transients (two separate transients in the same galaxy) or bright, slowly evolving pre-ZTF transients that were still visible at the start of ZTF. From this sample only SN~2017frh was found as a candidate with a late-time signal around 300 days after the first detection. Two more objects (SN ~2016cob and SN 2017fby) were found with a signal $>5$ years after the explosion, though a very thin but massive CSM shell is needed to explain the signal. The proximity of these SNe to their host galaxy's nuclei make it difficult to distinguish whether the signal is nuclear in nature or related to the SN.

The main issue that prevents \citet{JHT_DR2} and \citet{JHT_pre-ZTF} to be more conclusive than finding a few candidate objects is that the searches were done on archival data. Any signal that was found had already faded again by the time it was found, preventing follow-up photometry and/or spectroscopy to better characterise the signal. In this paper we adapt the pipeline from \citet{JHT_DR2} to work in (near) real time in an attempt to catch late-time signals as they occur and follow them up with different telescopes. In Section~\ref{data} we set up the pipeline and build the sample of objects to monitor. In Section~\ref{analysis} we describe the analysis steps that were taken when the pipeline flagged a possible active late-time signal, including follow-up observations if the signal looked promising. Section~\ref{results} describes the results of our monitoring campaign and is divided into three parts. Section~\ref{sec:faded_objects} describes the objects where a late-time signal is identified but could not be followed up on. Section~\ref{followups_section} describes the candidates that were followed up on, but the follow-up observations were inconsistent with a late-time signal due to CSM interaction. Section~\ref{sec:SN2020qxz} describes follow-up campaign and analysis of the late-time signal in SN~2020qxz. The results of our monitoring campaign are discussed in Section~\ref{discussion} and we conclude in Section~\ref{conclusions}. Throughout this paper, to convert between apparent and absolute magnitude we assume a flat $\Lambda$CDM cosmology with H$_0 = 67.7$ km s$^{-1}$ Mpc$^{-1}$ and $\Omega_\text{m} = 0.310$ \citep{Planck18VI}.

\section{Data}
\label{data}
To create a SN Ia sample to search for late-time CSM interaction in real time, we collect all spectroscopically classified SNe Ia in ZTF using the Fritz broker \citep{skyportal2019, Skyportal} and selected those objects that were first detected before 8 July 2023 (MJD 60133). Using this date cutoff ensures that all targets are over 100 days old at the start of the real-time search, and any new observations can be considered to be late-time observations of the SNe following the definition of \citet{JHT_DR2}. This gives us a sample of 6\,914 SNe Ia. Using the \textsc{fpbot} package \cite{fpbot}~\footnote{\url{https://github.com/simeonreusch/fpbot}} we construct the most up-to-date ZTF light curves for every object in our sample. ZTF uses difference imaging to remove constant sources from the science images and isolate the transient. Then, \textsc{fpbot} forces a PSF fit at a given location in the difference images to construct a light curve at that position spanning the entire survey.

As this sample contains all ZTF SNe Ia up until the first half of 2023, this naturally includes the entire ZTF SN Ia DR2, which have previously been examined for late-time signals in \citet{JHT_DR2}. 69 objects that were not available on Fritz when we made our sample have been excluded. However, as the latest data point used in \citet{JHT_DR2} is on 26 June 2022 (MJD 59756), we probe these SNe at later phases, and they could (have) become active after the time period examined by \citet{JHT_DR2}. Therefore, checking the same sample again could identify new late-time detections.

We aim to find late-time excesses while they are active, so the light curves have to be regularly updated with the latest observations. The ZTF survey is split into three parts: 40\% is a private partnership survey whose data we can access immediately after the observation and becomes publicly available after 18 months. 40\% is a public survey whose data is released every three months after a three-month proprietary period, resulting in a three to six-month delay before we can use this data. The last 20\% is private Caltech time, whose data is released publicly after 18 months.

We use the binning method from \citet{JHT_DR2} to search for faint signals, binning observations over 100 days after the SN peak together in bins of 25, 50, 75, and 100 days. (see Section~\ref{analysis}). The light curves do not have to be updated daily, as it will take several observations to push the detection limit deep enough to detect a faint signal in the latest bin. As the smallest bin size in the pipeline is 25 days, we aim to be able to create one additional bin each time an object is updated. We therefore split the sample into 28 lists, each containing 246 or 247 SNe, and update the objects in one of these each day. This results in every object in the sample being checked and updated once every four weeks.

\section{Analysis}
\label{analysis}
Each time the light curves of the SNe Ia in our sample are queried and updated, they are run through the analysis pipeline from \citet{JHT_DR2}. This pipeline takes all observations at a given sky position and after pre-processing (including applying a baseline correction to correct for systematic offsets across the light curve, see e.g. \citealt{Yao_baseline_corr, Miller_baseline_corr}) and finding the SN peak, it will bin all observations in each band that are over 100 days after the SN peak in bins that have a size of 25, 50, 75, or 100 days. By doing so, the noise in the data is reduced, allowing to probe objects that are up to nearly a magnitude below the noise limit of single observations ($m \approx20.5$ mag) at the price of a reduction in the time sensitivity. This method is able to push the detection limit until the magnitude limit of the used reference images become the dominant source of uncertainty, which is around $\overline{m} \approx21.8$ mag for ZTF \citep{ref_uncert}.

When the pipeline detects a possible late-time signal, whether currently or previously active, the object is rechecked manually to determine the possible origin of the late-time time signal and if it should be followed up. This includes checking the individual difference images for issues by using the Supernova Animation Program (\textsc{snap})\,\footnote{\url{https://github.com/JTerwel/SuperNova_Animation_Program}} that is first presented in \citet{JHT_DR2}. Possible explanations for the signal include, e.g., false positives due to a baseline offset, contamination from a possibly active host nucleus, known transients that are detectable over 100 days after the SN peak, or sibling transients at nearly the same sky position (see \citealt{JHT_DR2} and \citealt{JHT_pre-ZTF} for more details on these groups of objects). 

If the late-time signal is deemed real but has clearly faded again before it was found, all we can do is an analysis akin to that of \citet{JHT_DR2} and \citet{JHT_pre-ZTF}. However, if the signal is still active, we attempt to follow it up as quickly as possible using bigger telescopes to gain deep photometry (to confirm the late-time signal in a single observation) and spectroscopy (to search for spectral signature of the found signal). After all pre-existing light curves were generated, the real-time monitoring program started on 13 November 2023 (MJD 60261) and ran until 9 July 2024 (MJD 60500).

\subsection*{Follow-up observations and reduction}
Follow-up observations of promising late-time detections were carried out with the Optical System for Imaging and low-Intermediate-Resolution Integrated Spectroscopy (OSIRIS) instrument on the Gran Telescopio CANARIAS (GTC) and the Alhambra Faint Object Spectrograph and Camera (ALFOSC) on the Nordic Optical Telescope (NOT) at Observatorio Roque de Los Muchachos, La Palma. The NOT photometry is reduced using a custom pipeline in \textsc{python} using the \textsc{astropy} package. The GTC photometry is reduced with \textsc{iraf}. After the basic photometry reduction, difference imaging is done using \textsc{autophot} \citep{Autophot}\footnote{\url{https://github.com/Astro-Sean/autophot}}, using observations from Pan-STARRS \citep{Pan-STARRS1} or the Sloan Digital Sky Survey \citep[SDSS, ][]{SDSS-I-II, SDSS_DR4, SDSS_telescope} as reference images. 

The two most promising candidates were followed up with spectroscopy. GTC+OSIRIS spectra were obtained at the location of SN 2019zbq (discussed at the end of Section~\ref{followups_section}) at late times using grism R1000R and a 1.0\arcsec\ slit. They were reduced using a custom pipeline based on \textsc{pypeit} \citep{pypeit:joss_pub, pypeit:zenodo}. NOT+ALFOSC spectra obtained at the position of SN 2020qxz were obtained with grism 4 with a 1.3\arcsec\ slit on the 20, 30, and 31 March and 13, 14, and 15 April 2024. Spectra of the host galaxy of SN 2020qxz were obtained on 3 May and 19 June using a 1.0\arcsec\ slit. We used the WG345 order-blocking filter to remove second-order contamination of blue light in the red part of the host spectra. This was not used in the spectra of the transient as it was already very faint, and removing an extra $\sim$10\% of the light that hits the detector would decrease the SNR of the spectra. The spectra were reduced with~a data reduction pipeline on top of \textsc{pypeit}, available on GitHub \footnote{\url{https://github.com/steveschulze/NOT_DRP}}.

\begin{table*}
    \centering
    \caption{Objects with a detected late-time signal in the binned ZTF light curves.}
    \resizebox{\textwidth}{!}{
    \begin{tabular}{ccccccccccc}
        \hline
        \hline
        ZTF name & IAU name & RA & Dec & Type & $z$ & MJD$_\text{peak-SN}$ & MJD$_\text{disc-lt}$ & MJD$_\text{start-lt}$ & $\Delta t$ (day) & Comment\\
        (1) & (2) & (3) & (4) & (5) & (6) & (7) & (8) & (9) & (10) & (11)\\
        \hline
        ZTF18abtqevs & SN 2018grt & 00:04:36.30 & 51:59:37.63 & Ia-norm & $0.042\pm0.003$ & 58372 & 60279 & 59712 & 162 & \citet{JHT_DR2}\\
        ZTF19ablekwo & SN 2019mse & 00:15:21.36 & 46:44:08.64 & Ia-norm & $0.088\pm0.004$ & 58715 & 60283 & 59167 & 1170 $^{(2)}$ & \citet{JHT_DR2}\\
        ZTF20abjfufv & SN 2020tfc & 22:17:00.81 & 30:39:20.97 & Ia-norm & $0.031\pm0.001$ & 59116 & 60269 & 59671 & 804 $^{(3)}$ & \citet{JHT_DR2}\\
        ZTF20aaifyfx & SN 2020alm & 16:52:01.49 & 23:32:22.77 & Ia-norm & $0.06001\pm0.00001$ & 58873 & 60271 & 59624 & 826 $^{(3)}$ & \citet{JHT_DR2}\\
        ZTF22abgbbez & SN 2022uej & 07:19:57.72 & 46:00:59.00 & Ia & $0.03272\pm0.00008$ & 59839 & 60454 & 60168 & 212 & \ztfr\ detected only\\
        ZTF19acgonwy & SN 2019spg & 08:21:48.09 & -14:04:12.00 & Ia & $0.073\pm0.001$ & 58790 & 60353 & 58891 & 110 & Nuclear\\
        ZTF19abvvpoh & SN 2019pqn & 16:04:03.86 & 18:28:18.46 & Ia-norm & $0.03720\pm0.00001$ & 58745 & 60458 & 60436 & 62 $^{(3)}$ & Nuclear\\
        \hline
        ZTF20acpbbqf & SN 2020yvs & 10:24:22.78 & 41:42:25.97 & Ia-norm & $0.04453\pm0.00001$ & 59160 & 60268 & 59260 & 1127 & False positive\\
        ZTF20aazhzna & SN 2020jsa & 14:24:23.94 & 26:41:23.00 & Ia-norm & $0.03657\pm0.00001$ & 58992 & 60431 & 60315 & 72 & Nuclear\\
        ZTF18acaucsw & SN 2021nbt & 22:42:29.54 & 01:49:53.84 & Ia & $0.07\pm0.01$ & 59359 & 60283 & 59718 & 730 $^{(3)}$ & Nuclear / ANT\\
        ZTF19adcdgca & SN 2019zbq & 09:08:53.91 & 55:19:08.86 & Ia & $0.04678\pm0.00001$ & 58859 & 60261 & 59958 & 518 $^{(3)}$ & Nuclear / ANT \\
        ZTF20abqkbfx & SN 2020qxz & 18:04:00.24 & 74:00:50.07 & Ia-CSM & $0.0964\pm0.0004$ $^{(1)}$ & 59094 & 60388 & 60332 & 46 & Late-time line emission \\
        \hline
    \end{tabular}
    }
    \tablefoot{The table is divided into two parts, with the objects that were not followed up above and objects that were followed up below the division line. The first two columns give the ZTF and IAU names of the object, followed by the position, classified type, and redshift. The classifications are taken from the ZTF DR2 release \citep{DR2_spec_diversity, DR2_diversity}, except for SN 2022uej, whose classification comes from TNS. Column 7 gives the peak date of the SN, and column 8 gives the discovery date of the late-time signal. Column 9 gives the start of the first bin that is part of the late-time signal with the bin size that best shows the signal. Column 10 gives the duration of the late-time signal in rest frame days. The last column is a comment on the nature of the late-time signal.\\
    \tablefoottext{1}{$z$ value as derived from the classification spectrum, see Section~\ref{2020qxz_host_specs} for more details.}\\
    \tablefoottext{2}{Two periods of detections were identified.}\\
    \tablefoottext{3}{Active at the end of the monitoring period.}
    }
    \label{found_objs}
\end{table*}

Since a signal due to CSM interaction is not expected to evolve on the timescale of a few days \citep[e.g.,][]{2015cp}, we combined the spectra obtained in March 2024 of SN 2020qxz to gain a combined exposure time of 3 hours on the target. We also combined the spectra taken in April 2024 together to obtain a second spectrum with 2 hours and 40 minutes on the target. These will be called the `March' and `April' spectra of SN 2020qxz below (see Section \ref{sec:SN2020qxz}).

\section{Results}
\label{results}
The aim of this analysis is to identify in real-time SNe Ia with potential signatures of late-time interaction. Our pipeline identified 349 events that had a $5\sigma$ late-time detection, including ones that have been analysed in \citet{JHT_DR2}. 236 of these were ruled out as due to data issues (baseline offsets, etc.) and 101 due to known transients (tails, siblings), leaving 12 objects.

Compared to \citet{JHT_DR2} and \citet{JHT_pre-ZTF} the pipeline returns a slightly larger fraction of objects from our sample. In both \citet{JHT_DR2} and our sample, around two-thirds of the returned objects are false positives and around one-third are known objects, leaving the final 3-5\% to be investigated further. In \citet{JHT_pre-ZTF} around 72\% of the returned objects are false positives, and around 25\% are known objects, leaving a comparable 3\% of objects for deeper investigation. The main reason for this is because the \citet{JHT_pre-ZTF} sample  consists of objects first detected before the start of ZTF, leaving less opportunity to recover the declining tails of bright SNe, which is the group of previously known objects returned by the pipeline.

Table~\ref{found_objs} shows the 12 objects whose signal was detected in the binned ZTF light curves at $\geq100$ days after the SN peak. Four of the late-time detections were previously identified in \citet{JHT_DR2}. Even though their sample has been checked before, some objects might have become active after their search, but a visual inspection confirmed that this is not the case for the recovered objects. As their previous detections are already described in detail in \citet{JHT_DR2}, we do not consider them further here. In the following sections, we will firstly discuss the three new objects that have been found and were not previously identified but we could not follow up (Section \ref{sec:faded_objects}). After this we will discuss the five late-time detections that were followed up on and the interpretation of their data (Section \ref{followups_section}). 

\subsection{Faded objects}
\label{sec:faded_objects}
\begin{figure*}
    \centering
    \includegraphics[width=\textwidth]{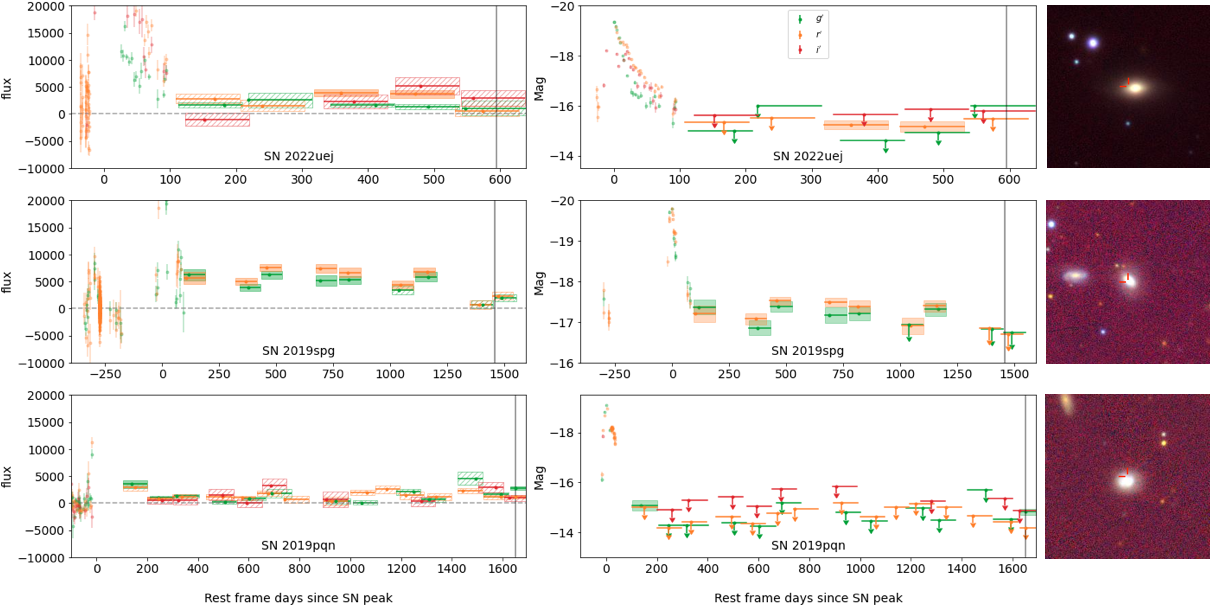}
    \caption{\textbf{Left:} Binned light curves in flux space in the SN rest frame of the three objects with a late time signal that we could not follow up on. The three ZTF bands \ztfg, \ztfr, and \ztfi\ are shown in green, orange, and red, respectively. Individual observations are shown before the binning starts. Bins are shown as coloured blocks showing the size of the bin, mean value, and $1\sigma$ uncertainty of the binned observations, with the point showing the mean observation date of the bin. Bins that are $>\ 5\sigma$ above zero flux are filled, and bins $\leq5\sigma$ from zero flux are dashed. \textbf{Middle:} Same plot as on the left but in absolute magnitude, corrected for MW extinction. $5\sigma$ detections in individual observations are shown before the binning starts, and the $5\sigma$ binned detections are shown in the same way as in the left plots. $5\sigma$ upper limits are shown with a downward arrow for the binned non-detections. The gray vertical line shows when the excess was first discovered. \textbf{Right:} PS1 cutouts centred on the SN location, which is marked in red.}
    \label{non_followup_lcs}
\end{figure*}

We found three new objects (SN 2022uej, SN 2020spg, SN 2019pqn) where a late-time detection was identified but we could not follow them up. In two cases, the late-time signal was recovered too late for follow-up observations, and in one case, we had no access to suitable telescope time to obtain constraining follow-up observations. The light curves of these objects are shown in Fig.~\ref{non_followup_lcs}.

\paragraph{SN 2022uej.}
A late-time \textit{r}-band signal in SN 2022uej was first picked up by our pipeline on 24 May 2024. It appears to have started 276 d in the rest frame before this and lasted over 200 d. There were observations in the \textit{g} and \textit{i} bands at similar epochs, but no detections at $>5\sigma$ were found. The date on which the \textit{r}-band detection was identified is a few days after a ZTF data release, which might have included some images from the public survey needed for a detectable signal. \textsc{snap} shows that the object is close to the imperfectly subtracted host nucleus, which causes some extra noise in the forced photometry light curve. However, the resulting dipole at the host location does not overlap with the SN location as it is oriented perpendicular to the direction of the SN, while there is a slight extra excess at the SN location 2.6\arcsec\ from the host. The identified signal is real in the data, but we were unable to follow up on the object because it had already faded by the time it was discovered.

\paragraph{SN 2019spg.}
The location of SN 2019spg has a relatively short, sparse baseline before the SN explosion. Immediately after the SN fades, the flux seems to settle at a plateau significantly above the pre-SN baseline. Due to it staying at a nearly constant flux level for much longer than the baseline, this could suggest that the baseline correction was incorrect and the actual baseline should be at the plateau level. With a separation of 0.5\arcsec, this object is also very close to the host galaxy nucleus. According to the AGN criterion presented in \citet{WISE_crit}, which uses data gathered with the \textit{Wide-field Infrared Survey Explorer} \citep[WISE, ][]{WISE}, the host galaxy is not an AGN. This does, however, not mean that the host cannot show small variability, as has been shown in \citet{JHT_DR2} and \citet{JHT_pre-ZTF}. Because of these reasons, we decided not to follow up on this late-time signal. After around 1\,000 days, the flux level dropped again, nearly to the pre-SN baseline level. This drop shows that the previous plateau was a real excess after all, although it was likely related to variability of the host nucleus, not the SN. Due to a combination of unfortunate timing of the excess and unfortunate sampling of the light curve, specifically the pre-SN light curve, these late-time detections of likely nuclear activity looked like a false positive until the signal faded and the original baseline was recovered.

Interestingly, the baseline correction was needed to recover the signal as the elevated plateau was originally at 0 flux. \textsc{snap} shows no excess in the difference images, but it does show a ghost (a negative imprint caused by over-subtraction) in the pre-SN region and after the binned signal has faded. This suggests that the excess is in the reference images and that the host dimmed between the time when the reference images were made and the SN explosion.

\paragraph{SN 2019pqn.}
A late-time signal at the position of SN 2019pqn was identified on 28 May 2024, only showing faint \ztfg\ band detections that had started 21 days earlier in the rest frame of the SN. This lasted at least 64 days and was active at the end of the monitoring period. The \ztfr- and \ztfi-bands only had upper limits at similar epochs. \textsc{snap} shows that the late-time signal lies between the SN and host nucleus locations (which are 1.9\arcsec\ apart) but seems to slightly favour the host nucleus location. When this excess was found, we did not have the opportunity to follow it up as we only had very low priority time at the NOT. Therefore, we only have the short period of binned photometry detections. Given that it was only detected in the \ztfg\ band, combined with the excess being slightly offset from the SN location, a nuclear transient is a good explanation of this late-time signal, even though the host nucleus is not an AGN according to its WISE colours.

\subsection{Follow-up results}
\label{followups_section}
\begin{figure*}
    \centering
    \includegraphics[width=\textwidth]{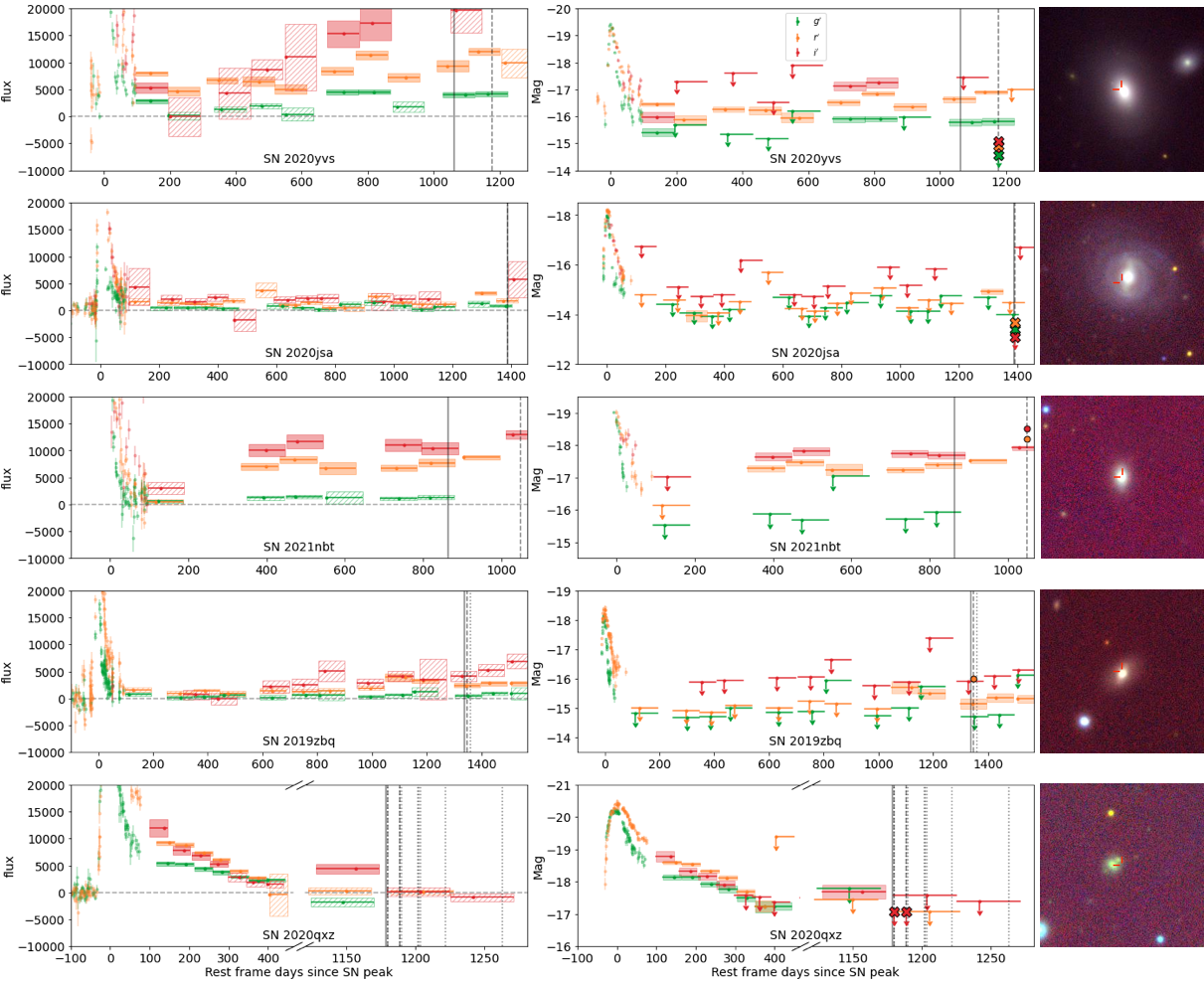}
    \caption{Same as Fig.~\ref{non_followup_lcs}, but for the objects for which follow-up observations were made. The grey vertical line shows when the excess was first discovered, and the grey dashed and dotted vertical lines show the photometric and spectroscopic follow-up observations, respectively. The coloured points at the follow-up dates show the follow-up detections and the crosses show the follow-up $5\sigma$ upper limits. Note that for SN 2020qxz the time period between the SN tail and late-time signal has been cut out to better show the detections and follow-up campaign.}
    \label{followup_lcs}
\end{figure*}

We found five objects with an active late-time signal for which we could obtain follow-up observations. SN 2019zbq was followed up with the GTC, while the others were followed up with the NOT. Below, we discuss each object individually, starting with the three for which only photometry was taken and then SN 2019zbq, for which we obtained photometry and spectroscopy. SN 2020qxz, for which more extensive follow-up was obtained, is discussed separately in Section \ref{sec:SN2020qxz}. The light curves of these objects are shown in Fig.~\ref{followup_lcs}. The details of their follow-up observations are presented in Table \ref{followups}.

\paragraph{SN 2020yvs.}
On 20 November 2023, SN 2020yvs was found to have elevated flux levels in all three ZTF bands that had started nearly 1\,000 d before, with the effect being stronger in the redder bands. The detected excess is clearly visible with \textsc{snap}, though at times it is merged with a residual dipole from imperfect subtraction of the host. As the SN is very close to the bright nucleus of the host galaxy (1.3\arcsec, host SDSS \ztfr\ mag = 13.56), we decided to keep a close eye on the development of the light curve to determine if it was more likely to be associated with the SN or not.

As the late-time detections grew stronger over time, we obtained confirmation through photometry with NOT/ALFOSC in the \textit{gri} bands. These observations were obtained on 19 March 2024, with ten exposures of 10 s each using the SDSS \ztfg\ztfr\ztfi\ filters that were stacked together for each band. The short exposures are needed to avoid saturating the host galaxy as image subtraction has to be performed to search for an $\sim 20$ mag excess in an $\sim 13.5$ mag environment. The image subtraction was done through \textsc{autophot} \citep{Autophot} using SDSS images as templates and was attempted using Pan-STARRS images as templates, though these proved to have a saturated host galaxy, making them unusable as templates.

After image subtraction, a small residual at the host nucleus was left, but no excess was found at the SN location. The host residual is the brightest source close to the location, having at most a $2\sigma$ detection at 21.6 mag in the \ztfi\ band. This is significantly below the detections found in the late-time ZTF light curve (brighter than mag 20 in both \ztfr\ and \ztfi). Therefore, we conclude that the binned detections are likely due to artefacts in the ZTF images after subtracting the bright host galaxy.

\paragraph{SN 2020jsa.}
The late-time signal in SN 2020jsa was found on 1 May 2024, showing a $21.2\pm0.1$ mag \ztfr-band excess that had begun 112 days earlier and was already fading at the time of discovery. The SN is located 3.0\arcsec\ from the host galaxy nucleus, and a previous faint detection in the binned late-time light curve can be seen around a year after the SN peak. While the host nucleus is relatively far from the SN, the difference images show a dipole with the positive side towards the SN after the host is subtracted. This would suggest that there are subtraction issues or that the host shows slight variability. However, according to the WISE criterion, the host galaxy is not an AGN. The object was observed with the NOT in \ztfg \ztfr \ztfi\ the next night, using short exposures to avoid saturation of the bright host and nearby stars. The resulting difference images found no excess at the SN location, providing a deeper limit than the binned ZTF \textit{r}-band light curve. However, a faint residual was found at the host nucleus location, showing that the excess is real but unlikely to be related to the SN. 

\paragraph{SN 2021nbt.}
The late-time signal in SN 2021nbt was first detected on 5 December 2023. The object is only 0.6\arcsec\ from the host nucleus, and its ZTF name is from 2018, which suggests that some host variability caused a first detection long before the SN exploded. This immediately raised suspicion that the detected late-time signal could be related to the host instead. Because of this, it took until 21 June 2024 before we decided to follow it up with the NOT. The signal started relatively shortly after the SN and persisted for several months, which could suggest it is connected to the SN in some way.

After difference imaging was performed on the \textit{ri}-band photometry obtained with the NOT (see Table \ref{followups}), a residual was recovered at \ztfr\ $= 19.458\pm0.040$ mag, \ztfi\ $= 19.123\pm0.062$ mag. This is somewhat brighter than the binned detections using ZTF observations taken around the same time, and a signal this bright would have been detectable in the individual ZTF observations. If the host nucleus varies slightly over time, however, this could cause different reference images to show the host at different brightness, which would result in different residuals after difference imaging. Unfortunately, there is no spectrum of the host galaxy, and since the signal is superimposed on the much brighter host, it is impossible to extract a spectrum of the excess without it being completely dominated by the host. The host is too small to attempt extracting a pure host spectrum away from the excess location, and without a host spectrum, the excess contribution cannot be isolated.

At the distance of the host, the magnitude of the excess observed by the NOT translates to an absolute magnitude of M $\sim -18$ mag. As can be seen in Fig.~\ref{followup_lcs}, the excess appears after a gap in the observations and persists for at least 780 days at a fairly constant magnitude. This is at the bright end of the potential ambiguous nuclear transient \citep[ANT,][]{2020ohl_Hinkle, wiseman_ztfants} population identified in \citet{JHT_pre-ZTF} but still consistent with the brightness of the comparison object they use. Due to its shape, brightness, and duration, an ANT that started somewhere between 200 and 400 days after the SN is a good explanation for this excess. However, we do not have spectroscopy to confirm since it was so faint relative to the host.

\paragraph{SN 2019zbq.}
On 13 November 2023, SN 2019zbq was noted to have an active late-time signal in the binned \ztfr-band at m $=21.5\pm0.2$ mag, with the signal already being present for nearly 300 days before and being recovered regardless of bin size or placement. The binned \ztfg-band did not rise from the baseline during this time, but the \ztfi\ band did, though the larger uncertainties in this band prevented the detections to reach a $5\sigma$ confidence level. The SN is close to the centre of its host galaxy, with an offset of $1.09\arcsec$.

Checking the \ztfr-band difference images with \textsc{snap} ruled out image defects, reduction or subtraction errors, or a sibling transient with a small sky separation from SN 2019zbq (within the uncertainty of the ZTF resolution) as potential explanations for the detections. The host galaxy has \textit{W1 -- W2} = $0.02\pm 0.04$ and  \textit{W2 -- W3} = $0.9\pm 0.5$, which is not an AGN according to the AGN criterion presented in \citet{WISE_crit}.

\begin{figure}
    \centering
    \includegraphics[width=\columnwidth]{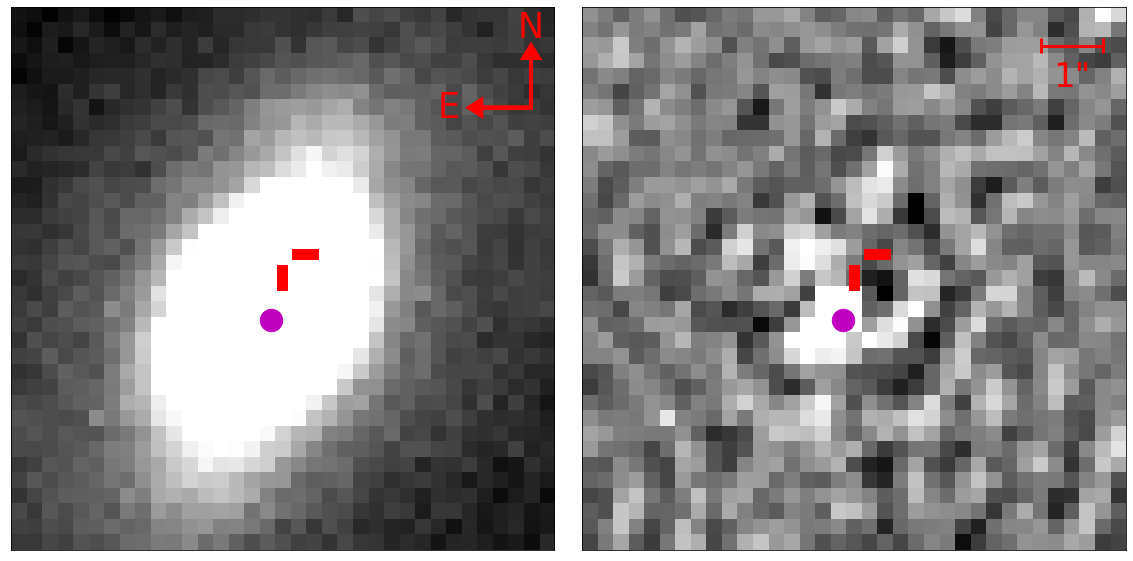}
    \caption{\textbf{Left:} \ztfr-band image of the location and host galaxy of SN~2019zbq, taken on 23 November 2023 with GTC+OSIRIS. The SN location is marked in red and the purple dot is the galaxy nucleus location. \textbf{Right:} Difference image of the left region after subtracting a Pan-STARRS template image. There is a residual visible at the host galaxy location. Note that the colour scaling is different for the two images to highlight the important sources.}
    \label{2019zbq_difim}
\end{figure}

We obtained a \ztfr-band image with the GTC and OSIRIS on 23 November 2023. Using \textsc{autophot} and reference images from Pan-STARRS, we recovered a m $=20.7\pm0.1$ mag source at the host nucleus location, which is clearly offset from the SN location, confirming the late-time excess is real but not related to the SN. The GTC image and the resulting difference image are shown in Fig.~\ref{2019zbq_difim}.

On 4 December 2023, we obtained a spectrum with the GTC and OSIRIS. Since the transient is located on top of the host, this spectrum contains light coming from both the host and the transient. The host is 4.5 mag brighter and dominates the spectrum heavily, meaning the host must be subtracted to recover any potential transient spectrum. Since a slit was used in obtaining the spectrum and the host galaxy is an extended object, we extract a pure host galaxy spectrum away from the transient location, and subtract it from the spectrum at the brightest part of the trace containing the transient. Assuming that the spectral features of the host are similar enough at these two locations, the main difference between the two, apart from total flux observed, will be the excess we are interested in.

Figure \ref{2019zbq_spec} shows the extracted host and excess spectra. Since the red edge of the detector has very low sensitivity, this can be used to scale the pure host spectrum to the combined host and excess spectrum, allowing us to take the difference and extract the pure excess spectrum. The pure excess spectrum is shown in the bottom panel. Integrating this spectrum over the \ztfr-band shows this excess to be $\sim20.2$ mag, which is slightly brighter than the photometry. The excess spectrum is quite noisy, but we can tentatively make out an emission feature that lines up with \HeliumI$_{\lambda7816}$. Overall, the spectrum looks red, which matches the binned ZTF data, as the \ztfg\ band does not rise from zero flux (and the \ztfi\ band rises the most but with a large scatter).

\label{2020qxz_host_specs}
\begin{table*}[]
    \centering
    \caption{Fits of the emission lines identified in the March 2024 spectrum at the position of SN 2020qxz.}
    \resizebox{\textwidth}{!}{
    \begin{tabular}{cccc|ccc}
        \hline
        \hline
        Central wavelength (\AA) & $\sigma$ (\AA) & $\sigma$ (km s$^{-1}$) & significance & Element & Rest wavelength (\AA) & Velocity offset (km s$^{-1}$)\\
        \hline
        4\,766.0 $\pm$ 1.1 & 12.7 $\pm$ 1.4 &  800 $\pm$ 88 $\pm\ 196_\text{inst}$ & 62 & \Hbeta & 4\,861.00 & $5\,920\pm70$\\
        8\,395.8 $\pm$ 0.6 & 9.3 $\pm$ 0.7 &  332 $\pm$ 25 $\pm\ 124_\text{inst}$ & 76 & \CaII\ $^{(1)}$ & 8\,542.00 & $5\,180\pm20$\\
        8\,423.6 $\pm$ 1.4 & 5.6 $\pm$ 1.5 &  199 $\pm$ 53 $\pm\ 124_\text{inst}$ & 14 & \NI & 8\,567.74 & $5\,090\pm50$\\
        8\,615.8 $\pm$ 1.3 &5.8 $\pm$ 1.5 & 202 $\pm$ 52 $\pm\ 122_\text{inst}$ & 17 & \KI\ $^{(2)}$ & 8\,763.96, 8\,767.05 & $5\,170\pm50$ \\
        \hline
    \end{tabular}
    }
    \tablefoot{The first three columns give the central wavelength and width (in \AA\ and km s$^{-1}$) of the fitted lines, we also give the instrumental velocity uncertainty at the emission line in the third column. The fourth column gives the significance of the detection. The last three columns give the most likely matched elements and emission lines, assuming some amount of blueshift velocity offset (given in the last column). The redshift uncertainty contribution to the velocity offset ($120$~km~s$^{-1}$) has not been included. All fits are done in the SN rest frame.\\
    \tablefoottext{1}{Only the middle line of the \CaII\ NIR triplet is visible, the other two lines are not present due to atmospheric and host contamination for the blue and red component, respectively.}\\
    \tablefoottext{2}{\KI\ is a doublet for which we use the mean to estimate the velocity offset.}
    }
    \label{fitres_2020qxz}
\end{table*}

The excess has an absolute magnitude M $=-16.0$ mag. This, together with the nuclear location and the sudden appearance of the transient followed by a relative stable plateau for over 500 days, is consistent with the faint ANT-like nuclear transients found in \citet{JHT_pre-ZTF}. Therefore, we conclude that a faint ANT-like transient provides a good explanation for the signal, although we caution that extracting the transient spectrum and removing the host contribution is not straightforward.

\subsection{SN 2020qxz}
\label{sec:SN2020qxz}
\begin{figure}
    \centering
    \includegraphics[width=\columnwidth]{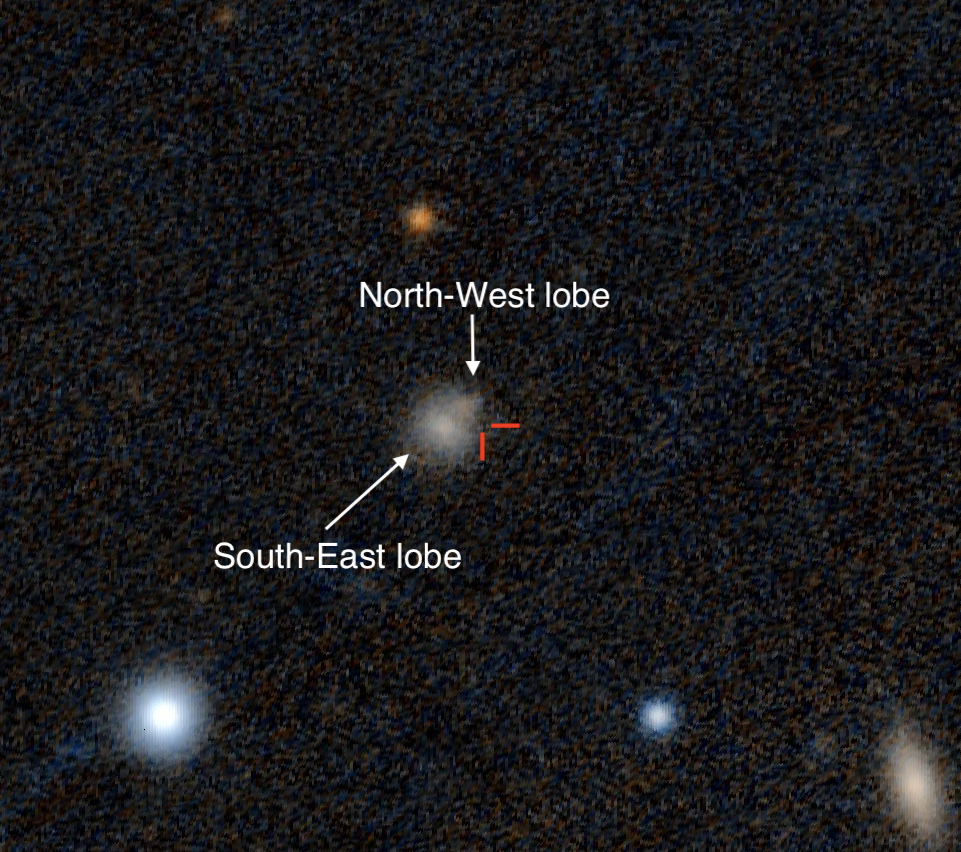}
    \caption{Pan-STARRS DR1 colour image of the location and host galaxy of SN 2020qxz. The SN location is marked in red, and the two host galaxy lobes are marked in white.}
    \label{2020qxz_loc}
\end{figure}

SN 2020qxz was classified as a SN Ia-CSM at early times \citep[][spectrum taken on 20-09-2020]{2020qxz_classif} with a prominent \Halpha\ line visible in the classification spectrum taken around the SN peak, which was used to obtain a redshift of $z=0.0964$ for the object. The slowly decaying tail is visible in the ZTF observations up to 1.2 years after the peak and an additional three months in the binned light curve before fading below the noise limit, resulting in its recovery in \citet{JHT_DR2}.

On 20 March 2024 (MJD 60389), at nearly 1\,180 d after the SN peak in the rest frame, this object was flagged as currently active due to an elevated flux level in the \ztfi-band in the last four weeks, with the detection being at m $=20.9\pm 0.2$ mag (see bottom row of Fig.~\ref{followup_lcs}). As can be seen in Fig.~\ref{2020qxz_loc}, the SN location is visually offset from the host galaxy, so we obtained follow-up observations with the NOT+ALFOSC on the same night. The resulting spectrum showed a prominent emission line around $\lambda=9\,210$ \AA\ in the observer frame ($\lambda=8\,400$ \AA\ in the SN rest frame). After this promising detection, we launched a campaign to take more observations (photometry and spectroscopy) with the NOT+ALFOSC over the following weeks.

\subsubsection{Late-time photometry}
Two epochs of \ztfi-band photometry were taken with the NOT+ALFOSC on 20 and 30 March 2024. Both were single 600 s exposures, which were reduced using the method described in Section \ref{analysis}. No residual flux above $5\sigma$ was found in either image, resulting in m = 21.5 mag upper limits. As can be seen in the bottom row of Fig. \ref{followup_lcs}, the two epochs of observations taken with the NOT (shown as crosses) were after the detection in the binned ZTF data. However, likely transient flux was detected in the spectra taken at the same epochs as this follow-up photometry. The most likely explanation for the photometric non-detections at the same epochs is that the emission lines seen in the spectra are no longer strong enough to cause a detectable increase in the broadband flux.

\subsubsection{Host spectroscopy}
The host galaxy of SN 2020qxz has two lobes, a big one on the South-East side and a smaller one on the North-West side (see Fig.~\ref{2020qxz_loc}). By orienting the slit along the axis of the galaxy, we obtained spectra of both lobes on the nights of 3 May and 19 June 2024 with the NOT+ALFOSC, resulting in a two-hour spectrum on the big (South-East) lobe and a one hour and 40 minutes spectrum on the small (North-West) lobe as one of the 20-minute exposures on 3 May 2024 was contaminated at the location of the small lobe and could not be used. 

The spectra of the big and small lobes shown at the bottom of Fig.~\ref{fig:2020qxz_spec} look nearly identical, and the galaxy emission and absorption lines discussed below are at the same position for both lobes. This shows that the host is more likely a single object at a single $z$ rather than two galaxies overlapping in the line of sight.

\begin{figure*}
    \sidecaption
    \includegraphics[width=12.97cm]{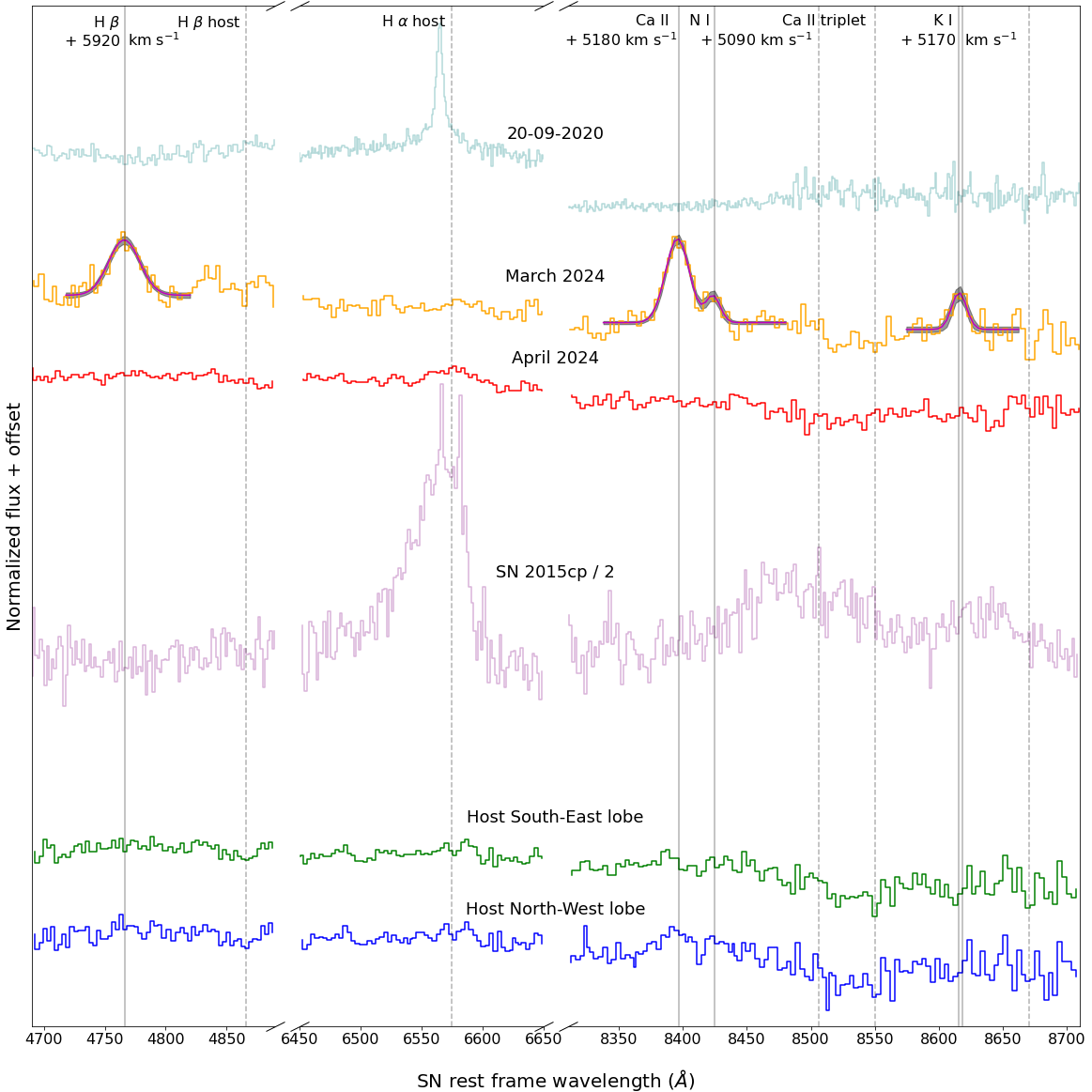}
    \caption{Combined late-time spectra of SN 2020qxz and its host in the SN rest frame, zoomed in on the regions containing the transient emission lines and the \Halpha\ region. The transient emission lines are marked with grey vertical lines, and the best-fit Gaussians are overlaid on top of the combined March spectrum with the $1\sigma$ uncertainties of the Gaussian fits shown in grey. The dashed vertical lines show the location of host galaxy lines (as well as \Hbeta\ to compare its location to the transient emission line) at the host redshift of $z=0.0975$. We also show the classification spectrum of SN 2020qzx taken around peak magnitude in light blue and the late-time spectrum obtained for SN 2015cp in magenta, scaled down by a factor of 2 for readability.}
    \label{2020qxz_spec_zoom}
\end{figure*}

Several lines can be seen in the host spectra. We identify weak lines that we associate with \Halpha\ emission, \SIIF\ emission, \CaII\ H\&K, and NIR triplet absorption, and \NaID\ and \MgI\ absorption. These galaxy lines give a galaxy redshift of $z=0.0975$, which is slightly larger than was found for the \Halpha\ emission seen in the maximum-light classification spectrum of SN 2020qxz as Ia-CSM (top spectrum in Fig.~\ref{fig:2020qxz_spec}). Assuming that the host galaxy is indeed at this larger $z$ and that the SN is associated with the host, the blueshifted \Halpha\ line in the classification spectrum of SN 2020qxz translates to the SN moving towards us with $\sim200$~km~s$^{-1}$ with respect to the host galaxy. This is well within the dispersion of a typical galaxy, which could explain the SN velocity offset from its host. The SN could also have occurred in an unseen dwarf satellite. The different $z$ value would also slightly change the distance modulus, resulting in all absolute magnitudes being increased by $\sim0.03$ mag. Unless specified otherwise, we use the SN rest frame at $z=0.0964$ while further discussing this object.

\subsubsection{Late-time spectroscopy at the SN location}
\label{2020qxz_spec}
We obtained spectra at several epochs at the position of SN 2020qxz in March and April 2024, which were combined into a `March' spectrum and `April' spectrum (see Section \ref{analysis} for further details). Fig~\ref{2020qxz_spec_zoom} shows the most interesting regions of the spectra. The full spectra are shown in Fig.~\ref{fig:2020qxz_spec}, and in Figs.~\ref{fig:2020qxz_spec_all_lines} and~\ref{2020qxz_spec_zoom_all_lines} we show the spectra from the individual nights that are combined to make these two spectra. In the March spectrum, we identify four prominent emission lines and zero absorption lines that are not associated with the host galaxy. Both the March and April spectra show some signs of slight host contamination, but the four emission lines are only seen in the March data, not in the April spectrum nor in the host spectra obtained in May and June 2024. This suggests that these features are transient in nature, and are decaying between the two periods of observations. These emission lines are seen at $4\,766.0\pm1.1$ \AA, $8\,395.8\pm0.6$ \AA,  $8\,423.6\pm1.4$ \AA, and $8\,615.8\pm1.3$ \AA\ in the rest frame of the SN and are marked in Fig.~\ref{2020qxz_spec_zoom} and Fig.~\ref{fig:2020qxz_spec}. 

We fitted these four emission lines with Gaussian profiles to measure their strengths and widths, the results of which are given in Table~\ref{fitres_2020qxz}. The fits to the four lines are shown in Figs.~\ref{2020qxz_spec_zoom} and~\ref{fig:2020qxz_spec}. Three of the lines (the three at $\sim$8400 - 8600 \AA) have a similar dispersion, suggesting that they are from the same source. However, the bluest line has a much higher dispersion of 800 $\pm88$  km s$^{-1}$, which is incompatible with the other three. 

Identification of the origin of these emission lines is not straightforward. As SN 2020qxz is a known SN Ia-CSM that showed \Halpha\ emission near peak brightness, it is natural to assume that at least one of the four lines in the March 2024 spectrum is due to H emission. Indeed, the line at $4\,776$ \AA~can be interpreted as \Hbeta\ emission, though this requires it to be blueshifted by $5\,920\pm70$ km s$^{-1}$. Offset emission lines have been found before at up to 10\,000 km s$^{-1}$ for the rapidly declining Ca-rich event, SN 2019bkc, see e.g. \cite{2019bkc_Chen} and \cite{2019bkc_Prentice}. However, the significant difference in the hydrogen velocity between the early and late-time detections suggests that this material has been accelerated in the time between detections or is coming from a different location.

SN ejecta sweep up the CSM they are interacting with, so some acceleration can be expected. This has been seen before in some SNe IIn, where aspherical, possibly clumpy CSM has been accelerated to several thousand km s$^{-1}$ resulting in wide boxy or multi-peaked emission lines \citep{1998S_aspherical_CSM, 1998S_late-time, PTF11iqb}. The other issue is that where there is \Hbeta\ emission, one would expect \Halpha\ emission as well. However, there is no sign of any emission line at the position of \Halpha\ in the March or April spectra. There is a small bump that is consistent with being due to host \Halpha\ contamination, but nothing is seen with the same offset as \Hbeta. 

The other three transient emission lines at 8\,395.8 \AA,  8\,423.6 \AA, and 8\,615.8 \AA\ in the rest frame of the SN  have their own identification challenges. First of all, these are in the same region as the host galaxy \CaII\ NIR absorption. For the transient emission lines to be visible, this dip in the background spectrum has to be overcome first. Three emission lines are found, but their relative positions are different from the \CaII\ NIR triplet, meaning that not all lines can be explained by them. However, if we shift the middle line of this triplet to match with the bluest transient line, the blue line in the \CaII\ NIR triplet overlaps with a skyline while the red line is in the middle of the dip mentioned before. This could explain why only one of the three components is visible. The required blueshift is $5\,180\pm20$ km s$^{-1}$, significantly lower than the shift required for \Hbeta\ of $5\,920$ km s$^{-1}$. A shift of $\sim5\,150$ km s$^{-1}$ allows the remaining two transient lines to be matched to a \NI\ line and \KI\ doublet with the same velocity, resulting in a possible identification of three offset emission lines with similar widths. For the emission line matching with the \KI\ doublet we use the mean wavelength.

Neither of these elements are expected to have a strong presence in the ejecta of SNe Ia, nor have these lines been seen in other CSM interacting SNe. \citet{2014J_KI} did, however, find time-varying \KI\ absorption lines in SN 2014J around its peak, which they associated with photoionization of distant CSM, although the inferred distance is larger than what the ejecta could travel in a couple of years. It might also be possible to obtain a nitrogen-enhanced CSM if the donor star is an evolved star massive enough to burn hydrogen through the CNO cycle but small enough for the primary to have ended up as a WD. Dredge-ups could bring fusion products to the surface, including nitrogen, which is then lost from the system to form the CSM. \citet{N_lines_SECCSN} make a similar argument for the production of [\NII] lines in stripped envelope CCSNe.

The late-time CSM in SN 2015cp was primarily \Halpha\ dominated \citep{2015cp}, which is absent in SN 2020qxz (see comparison in Figs.~\ref{2020qxz_spec_zoom} and~\ref{fig:2020qxz_spec}). At the same time, three of the four emission lines found in SN 2020qxz are absent in 2015cp. There is a tentative detection of the \CaII\ NIR triplet in SN 2015cp, but unlike in SN 2020qxz, all three components are observed. They only require a velocity offset of $\sim1\,400$ km s$^{-1}$, though with a FWHM of $\sim3\,000$ km s$^{-1}$ the \CaII\ lines in SN 2015cp are much broader than any of the transient emission lines found in SN 2020qxz.

\section{Discussion}
\label{discussion}
In this study we monitored a sample of 6\,914 SNe Ia discovered by ZTF for signs of late-time rebrightening due to potential CSM interaction. We created forced photometry light curves at the location of each SN for the full duration of the survey and updated these every four weeks with the latest observations in an attempt to find late-time signals in real time and follow them up with further observations. We found 12 objects with late-time rebrightening that could not be explained by image defects, known contamination of host nuclei, or known sibling transients. Four of these were known from \citet{JHT_DR2} and discussed there. Their rediscovery here did not lead to any new insights. In three objects, the signal was found when follow-up was not possible. We were able to follow up the signals in the other five objects, using the NOT for four and the GTC for one. Out of these, there was one false positive, three likely nuclear transients unrelated to the SN, and a confirmed late-time signal in SN 2020qxz.

\subsection{Detection challenges}
As was shown in \citet{JHT_DR2} and \citet{JHT_pre-ZTF}, late-time signals in SNe Ia are rare. Out of nearly 7\,000 objects, we found eight objects with a late-time signal detected somewhere in the 240 days we were actively monitoring them and an additional four objects whose late-time signals were already discussed in \citet{JHT_DR2}. During this time many other objects were flagged but turned out to be false positives, bright long-lived SNe, or sibling transients. SNe near their host nucleus are especially prone to give false detections.

Most objects for which a late-time signal was flagged in Table~\ref{found_objs}, \citet{JHT_DR2}, and \citet{JHT_pre-ZTF} are close to the host nucleus. These regions are difficult to handle with difference imaging, often leading to residuals that can be mistaken for a weak signal by forced photometry. Even if the excess is real, the complicated centres of galaxies leave plenty of opportunity for alternative explanations, including ones involving the central supermassive black hole such as AGN, tidal disruption events and other sources of nuclear variability, e.g., ANTs \citep{2020ohl_Hinkle}. One could make a cut on the host separation to ensure that contamination from nuclear events is removed, but we did not do this here as it would introduce a bias in the SN environments in our sample. Such a cut is also dependent on the pixel size of the observations, which means that more distant SNe require larger host separations in order to not be removed by the cut. We decided to instead be as inclusive as possible with the knowledge of the contamination this leads to in our results.

Even when searching for faint late-time signals in real time, the fact that our method requires several observations to bin together in order to reach a deeper detection limit results in our method still being slightly behind current events. This makes it crucial to follow up quickly on newly found signals with bigger telescopes and better instruments that are able to measure such faint signals, especially as sometimes the late-time signals have been relatively short-lived and were already fading again by the time they were found. Without our ability to follow up quickly, the transient emission lines in SN 2020qxz (see Section \ref{sec:SN2020qxz}) would likely have been missed.

\subsection{Follow-up campaigns}
For five objects that showed late-time signals we were able to get follow-up photometry, on which difference imaging and forced photometry was performed using \textsc{autophot}. In one case (close to the position of SN 2020yvs), the follow-up observations provided deeper non-detections than the late-time signal that was found in the binned ZTF observations. The host galaxy of SN 2020yvs is very bright, which likely results in small artefacts in the difference images that are misidentified as the late-time signal. In three other cases (close to the positions of SNe 2020jsa, 2021nbt and 2019zbq), a residual was detected at the host nucleus location, showing that for these objects the recovered late-time signal was unrelated to the SN. 

The host galaxy of SN 2019zbq was confirmed to have an active nuclear transient and was followed up with spectroscopy. However, since the host galaxy is 4.5 mag brighter than the transient, it dominates the spectrum heavily, and the only way to extract the transient spectrum is by subtracting the host contribution. This requires the existence of a pre-transient host spectrum \citep[from e.g. SDSS, as was done for the late-time signal in SN 2020alm in][]{JHT_DR2} or a host spectrum without contribution from the transient extracted from the same observation as the spectrum containing the transient, as was done for SN 2019zbq in Sect.~\ref{followups_section}. This is a tricky process and can introduce a lot of uncertainty. For this reason, we did not attempt to get a spectrum of the confirmed excess in the host of SN 2021nbt, as there was already enough evidence of the late-time excess being unrelated to the SN. Integrating the flux of the SN 2019zbq host excess spectrum over the \ztfr-band matches the measured \ztfr-band brightness, showing that the excess extraction has worked. 

The excess spectrum in the SN 2019zbq host looks quite red but has no distinctive emission or absorption features. This is similar to the ANT AT 2020ohl \citep{2020ohl_Hinkle}, although AT 2020ohl is much bluer and brighter. The ANTs presented in \citet{wiseman_ztfants} are much closer to our excess in colour but again much brighter, and they all show Balmer emission lines. The small peaks that do stand out in our excess spectrum do not match any set of elements at a single velocity offset. The best we can say about it is that this is indeed some sort of ambiguous nuclear transient.

\subsection{The late-time signal in SN 2020qxz}
SN 2020qxz stands out from the other objects in the sample for several reasons. First of all, it is a SN Ia-CSM with detected interaction around its peak. It is also visually separated from its host nucleus, which makes it easier for the excess to be interpreted as related to the SN instead of nuclear activity, and it is also easier to get a spectrum of the excess without much contribution from the host galaxy. We followed this excess up over three months and obtained nearly six hours of spectroscopic data of the transient, which we combined into a March spectrum and an April spectrum. We also observed the host galaxy for one hour and 40 minutes. We also got a spectrum of the two lobes of the host galaxy to confirm that they are not two galaxies overlapping in the line of sight and finding a host redshift of $z=0.0975$. This is slightly larger than the estimated $z=0.0964$ from the \Halpha\ line in the SN peak spectrum, and gives the SN a velocity of $\sim200$~km~s$^{-1}$ with respect to the host galaxy. The SN is 3.4\arcsec\ from the centre of the South-East lobe, which translates to $\sim7$~kpc at the host redshift. This means that the SN velocity can be interpreted as its rotation velocity around the host.

The March spectrum showed four emission lines that are not in the host spectrum, and these had disappeared a few weeks later in the April spectrum. The transient nature of these lines shows that they are associated with the \ztfi-band detection in the binned ZTF light curve. As the follow-up was performed when the late-time signal was already fading again, both the photometric follow-up and the binned observations resulted in non-detection during the follow-up campaign. The remaining transient emission lines are too faint to be picked up in the broadband photometry and are only found during the long spectroscopic observations. This suggests that we only managed to obtain the tail end of the identified signal with spectroscopy. 

The only way to match the transient emission lines to elements that can be expected in such an environment is to allow for large velocity offsets. The three lines around 8500~\AA\ all require a velocity of $\sim5\,150$ km s$^{-1}$ when matching them with one of the lines of the \CaII\ NIR triplet (the other two are likely contaminated by the host lines, see Section \ref{2020qxz_host_specs}), \NI, and \KI\ emission lines. Fitting the lines with Gaussians shows that they have similar widths, suggesting that the elements producing these lines are in the same location. The bluest transient line, however, is best matched by \Hbeta\ with an offset of $5\,920\pm70$ km s$^{-1}$, and it is also broader than the other three lines. This suggests that, if this line identification is correct, the H-rich material is in a different and faster moving cloud than the N, K, and Ca. It is important to note that although we did not use an order-blocking filter for these observations, none of the emission lines match a second-order diffraction of another emission line, showing that they are four different emission lines.

One problem with the interpretation of the emission line as being \Hbeta\ is that in most cases, when \Hbeta\ is detected, one would also expect to detect \Halpha\ with at least the same strength. This is not the case here, nor is there any sign of an absorption feature at the expected location of \Halpha\ that could completely wipe out a strong, relatively broad emission line like this. Late-time observations of SN 1987A have allowed for detailed studies of the interaction between the ejecta and circumstellar material, with many emission lines having been found contributing to the overall luminosity \citep[see e.g.][]{87A_CSM_emission_1st_years, 87A_late_emission_lines, 87A_CSM_abundances}. The main observed emission line near 4766 \AA\ besides \Hbeta\ is \HeII$_{\lambda4\,686}$. Matching our bluest transient emission line to this would require a similar velocity offset as for \Hbeta, but it would be redshifted instead of blueshifted. On top of that, there are no other lines that can simultaneously explain the other three emission lines in the late-time spectra of SN 2020qxz. \MgI$_{\lambda4\,740}$ is also an interesting candidate for the line at 4766~\AA\ with a smaller required velocity, but it suffers from the same problem of having no candidates to fit the other three lines at a similar velocity offset. Therefore, we find that is \Hbeta\ the best match despite its shortcomings.

The classification of SN 2020qxz as a SN Ia-CSM is due to the \Halpha\ signal around the SN peak from interaction with CSM relatively close to the progenitor system. The late-time signal suggests a second region of CSM further from the progenitor system. Assuming that the inner CSM has a shell or torus-like structure, the interaction signature has the same total redshift as the SN. The late-time transient lines require a significant velocity offset, and their Gaussian profiles were much wider than that of the \Halpha\ line. This shows that the outer CSM is moving much faster and has much more of a patchy structure. Velocity offsets of a few thousand km s$^{-1}$ have been seen before in nebular emission lines of SNe Ia \citep{Maeda_exp_asymetry, Maguire_opt_NIR}. $>5\,000$ km s$^{-1}$ is fast but not unprecedented, as the \CaII\ lines in the very rapidly evolving SN 2019bkc have been measured to be up to $10\,000$ km s$^{-1}$ \citep{2019bkc_Chen, 2019bkc_Prentice}.

CSM with velocity offsets in this regime has been seen before in some SNe IIn such as SN 1998S \citep[$\sim4\,500$~km~s$^{-1}$][]{1998S_aspherical_CSM, 1998S_late-time} and PTF11iqb \citep[$\sim3\,000$~km~s$^{-1}$][]{PTF11iqb}. It has been interpreted as material that has been accelerated as it has been swept up. If the CSM is clumpy or disk-like, this can result in a multi-peaked emission line where the redshifted peak is weakened as it comes from the receding end of the SN and has had to travel through dust-forming SN material. In the case of SN 2020qxz, the late-time signal could be interpreted as the ejecta running into a shell of H-rich CSM around peak light, resulting in the \Halpha\ line in the classification spectrum. As the CSM is swept up, it is accelerated to at least $\sim6\,000$ km s$^{-1}$ before a part of the accelerated ejecta crashes into a cloud of CSM, resulting in the interaction signal that was observed in 2024. Only the expanding part of the ejecta with accelerated CSM that interacts with this CSM cloud would be emitting light, explaining the offset. However, the viability of such a scenario needs to be tested with modelling.

\section{Conclusions}
\label{conclusions}
In this work, we have presented a real-time search of late-time signals in 6\,914 Type Ia SNe in ZTF, using the latest observations to create the most up-to-date light curves for the objects in our sample. By binning the latest observations we were able to go up to nearly one magnitude beyond the single observation limit, which allowed us to find eight new faint signals during our 240-day monitoring effort. Using the NOT and GTC, three of these objects were subsequently followed up with photometry, and two objects were followed up with photometry and spectroscopy. Our main conclusions are:

\begin{enumerate}
    \item The requirement to bin observations to get deeper magnitude limits makes even the most real-time strategy lag slightly behind current events. In some cases this resulted in the discovery of a (short-lived) late-time signal while it was already declining again. Quick follow-up is required, but only the tail end of the signal can likely be observed.
    \item Late-time signals are rare, and detections of late-time signals that are associated with SNe are contaminated with detections of unassociated objects in the same line of sight. This is more prevalent for SNe close to the their host's nucleus.
    \item In five objects, the late-time detections have been determined to be due to nuclear activity. One of these was not followed up on as it was initially dismissed as a false positive until it faded. This shows the importance of a good baseline when searching for faint signals. The three objects that were followed up showed an excess at the location of the host nucleus, clearly separated from the SN position due to the better resolution.
    \item The excess in the SN 2019zbq host was followed up spectroscopically and showed both similarities and differences to previously studied ANTs, with the main difference being the absolute magnitude of the signal. Since the excess is 4.5 mag fainter than the host nucleus, extracting its spectrum required careful subtraction of the host signal, which complicates studying faint ANTs like these.
    \item The late-time signal in SN 2020qxz stood out from the others, as this is a SN Ia-CSM at $\sim7$ kpc from the host nucleus, making it significantly easier to follow up. While the signal was short-lived, we were able to follow it up spectroscopically, finding four transient emission lines that are best matched by four different elements at two different, highly blueshifted velocities ($5\,150 - 5\,920$ km s$^{-1}$). This suggests that the progenitor system created at least two regions of CSM: one nearby (resulting in the initial Ia-CSM classification) and one further away that caused in the late-time signal.
\end{enumerate}

Faint late-time signals in SNe Ia due to interaction with distant CSM are rare, and even with techniques like binning observations to reach the deepest detection limits, only the brightest and strongest signals can be detected with current surveys. The upcoming Vera C.~Rubin Observatory's Legacy Survey of Space and Time \cite[LSST;][]{LSST} will be able to detect objects several magnitudes deeper than ZTF, allowing the detection of fainter objects in a larger volume. Monitoring existing SN samples like those found by ZTF with LSST will be the best way to find faint late-time signals, which could then be followed up spectroscopically to find the chemical composition of the source of the signal. Constraining the properties of these rare events will allow us to probe the physics that govern SN Ia explosions more thoroughly and gain new insights to help explain them.

\begin{acknowledgements}
    JHT and KM acknowledge support from EU H2020 ERC grant no. 758638. SJB acknowledges their support by the European Research Council (ERC) under the European Union’s Horizon Europe research and innovation programme (grant agreement No. 10104229 - TransPIre). L.G. acknowledges financial support from AGAUR, CSIC, MCIN and AEI 10.13039/501100011033 under projects PID2023-151307NB-I00, PIE 20215AT016, CEX2020-001058-M, ILINK23001, COOPB2304, and 2021-SGR-01270. T.E.M.B. and U.B are funded by Horizon Europe ERC grant no. 101125877. Y.-L.K. was supported by the Lee Wonchul Fellowship, funded through the BK21 Fostering Outstanding Universities for Research (FOUR) Program (grant No. 4120200513819) and the National Research Foundation of Korea to the Center for Galaxy Evolution Research (RS-2022-NR070872, RS-2022-NR070525).
    
    Based on observations obtained with the Samuel Oschin Telescope 48-inch and the 60-inch Telescope at the Palomar Observatory as part of the Zwicky Transient Facility project. ZTF is supported by the National Science Foundation under Grants No. AST-1440341 and AST-2034437 and a collaboration including current partners Caltech, IPAC, the Oskar Klein Center at Stockholm University, the University of Maryland, University of California, Berkeley, the University of Wisconsin at Milwaukee, University of Warwick, Ruhr University, Cornell University, Northwestern University and Drexel University. Operations are conducted by COO, IPAC, and UW.

    Based on observations made with the Nordic Optical Telescope, owned in collaboration by the University of Turku and Aarhus University, and operated jointly by Aarhus University, the University of Turku and the University of Oslo, representing Denmark, Finland and Norway, the University of Iceland and Stockholm University at the Observatorio del Roque de los Muchachos, La Palma, Spain, of the Instituto de Astrofisica de Canarias. The data presented here were obtained [in part] with ALFOSC, which is provided by the Instituto de Astrofisica de Andalucia (IAA) under a joint agreement with the University of Copenhagen and NOT.
    
    Based on observations made with the Gran Telescopio Canarias (GTC), installed at the Spanish Observatorio del Roque de los Muchachos of the Instituto de Astrofísica de Canarias, on the island of La Palma. This work is (partly) based on data obtained with the instrument OSIRIS, built by a Consortium led by the Instituto de Astrofísica de Canarias in collaboration with the Instituto de Astronomía of the Universidad Autónoma de México. OSIRIS was funded by GRANTECAN and the National Plan of Astronomy and Astrophysics of the Spanish Government.
    
    This research made use of {\ttfamily PypeIt},\footnote{\url{https://pypeit.readthedocs.io/en/latest/}} a Python package for semi-automated reduction of astronomical slit-based spectroscopy \citep{pypeit:joss_pub, pypeit:zenodo}.

    The Pan-STARRS1 Surveys (PS1) and the PS1 public science archive have been made possible through contributions by the Institute for Astronomy, the University of Hawaii, the Pan-STARRS Project Office, the Max-Planck Society and its participating institutes, the Max Planck Institute for Astronomy, Heidelberg and the Max Planck Institute for Extraterrestrial Physics, Garching, The Johns Hopkins University, Durham University, the University of Edinburgh, the Queen's University Belfast, the Harvard-Smithsonian Center for Astrophysics, the Las Cumbres Observatory Global Telescope Network Incorporated, the National Central University of Taiwan, the Space Telescope Science Institute, the National Aeronautics and Space Administration under Grant No. NNX08AR22G issued through the Planetary Science Division of the NASA Science Mission Directorate, the National Science Foundation Grant No. AST-1238877, the University of Maryland, Eotvos Lorand University (ELTE), the Los Alamos National Laboratory, and the Gordon and Betty Moore Foundation.
    
    Funding for the SDSS and SDSS-II has been provided by the Alfred P. Sloan Foundation, the Participating Institutions, the National Science Foundation, the U.S. Department of Energy, the National Aeronautics and Space Administration, the Japanese Monbukagakusho, the Max Planck Society, and the Higher Education Funding Council for England. The SDSS Web Site is http://www.sdss.org/.

    The Gordon and Betty Moore Foundation, through both the Data-Driven Investigator Program and a dedicated grant, provided critical funding for SkyPortal.
    
    The ztfquery code was funded by the European Research Council (ERC) under the European Union's Horizon 2020 research and innovation programme (grant agreement n°759194 - USNAC, PI: Rigault).
    
\end{acknowledgements}

\section*{Data Availability}
The ZTF light curves were generated using \textsc{fpbot}\footnote{\url{https://github.com/simeonreusch/fpbot}}. The binning program can be found at \url{https://github.com/JTerwel/real-time_interaction_search} and includes the full sample of objects used in this paper. Also, \textsc{snap} can be found at \url{https://github.com/JTerwel/SuperNova_Animation_Program}. The late-time host and host+excess spectra of SN 2019zbq, the individual late-time spectra of SN 2020qxz and the combined SN 2020qxz host spectra are available on WISeREP\footnote{\url{https://www.wiserep.org}} \citep{wiserep}. All other follow-up observations are available upon request to the author.

\bibliographystyle{aa}
\bibliography{main}

\begin{appendix}
\onecolumn
\section{Log of follow-up observations}
\begin{table*}[h!]
    \centering
    \caption{Follow-up observations of transients with late-time signals.}
    \resizebox{\textwidth}{!}{
    \begin{tabular}{cccccccc}
        \hline
        \hline
        Name & Tel. / Inst. & Mode & Setup & MJD & Exp. time (s) & Seeing (\arcsec) & Comment\\
        \hline
        SN 2020yvs & NOT / ALFOSC & phot & \ztfg\ztfr\ztfi & 60388 & 10 x 10 & 0.9 & \ztfg\ > 22.2, \ztfr\ > 21.8, \ztfi\ > 21.6 mag \\
        SN 2020jsa & NOT / ALFOSC & phot & \ztfg\ztfr\ztfi & 60432 & 10 x 30 & 1.0 & \ztfg\ > 22.76, \ztfr\ > 22.49, \ztfi\ > 23.05 mag, host residual\\
        SN 2021nbt & NOT / ALFOSC & phot & \ztfr\ztfi & 60482 & 4 x 30 & 0.9 & \ztfr\ $= 19.458\pm0.040$, \ztfi\ $= 19.123\pm0.062$ mag\\
        \multirow{2}{*}{SN 2019zbq} & \multirow{2}{*}{GTC / OSIRIS} & phot & \ztfr & 60271 & 1 x 10 & 1.2 & \ztfr\ = $20.7 \pm 0.1$ mag\\
        & & spec & Grism R1000R, 1.0\arcsec slit & 60282 & 3 x 1200 & 1.2 & Contains host \& excess spectra\\
        \hline
        \multirow{10}{*}{SN 2020qxz} & \multirow{10}{*}{NOT / ALFOSC} & phot & \ztfi & 60389 & 1 x 600 & 1.7 & \ztfi\ > 21.5 mag\\
        & & spec & Grism \#4, 1.3\arcsec slit & 60389 & 3 x 1200 & 1.7 & Part of March spectrum\\
        & & phot & \ztfi & 60399 & 1 x 600 & 1.4 & \ztfi\ > 21.5 mag\\
        & & spec & Grism \#4, 1.3\arcsec slit & 60399 & 3 x 1200 & 1.4 & Part of March spectrum\\
        & & spec & Grism \#4, 1.3\arcsec slit & 60400 & 3 x 1200 & 0.8 & Part of March spectrum\\
        & & spec & Grism \#4, 1.3\arcsec slit & 60413 & 3 x 1200 & 1.5 & Part of April spectrum\\
        & & spec & Grism \#4, 1.3\arcsec slit & 60414 & 3 x 1200 & 1.7 & Part of April spectrum\\
        & & spec & Grism \#4, 1.3\arcsec slit & 60415 & 2 x 1200 & 1.3 & Part of April spectrum\\
        & & spec & Grism \#4, 1.0\arcsec slit $^{(*)}$ & 60433 & 3 x 1200 & 1.4 & Part of host spectrum\\
        & & spec & Grism \#4, 1.0\arcsec slit $^{(*)}$ & 60480 & 3 x 1200 & 0.7 & Part of host spectrum\\
        \hline
    \end{tabular}
    }    
    \tablefoot{The exposure time is noted as the number of exposures x exposure time of a single exposure. Photometric observations taken in multiple bands with the same exposures are mentioned in the same row.\\
    \tablefoottext{*}{A WG345 order blocking filter was used to remove second order contamination in the red part of the spectrum.}
    }
    \label{followups}
\end{table*}

\section{Late-time spectrum of SN 2019zbq}
\begin{figure*}[h!]
    \centering
    \includegraphics[width=\textwidth]{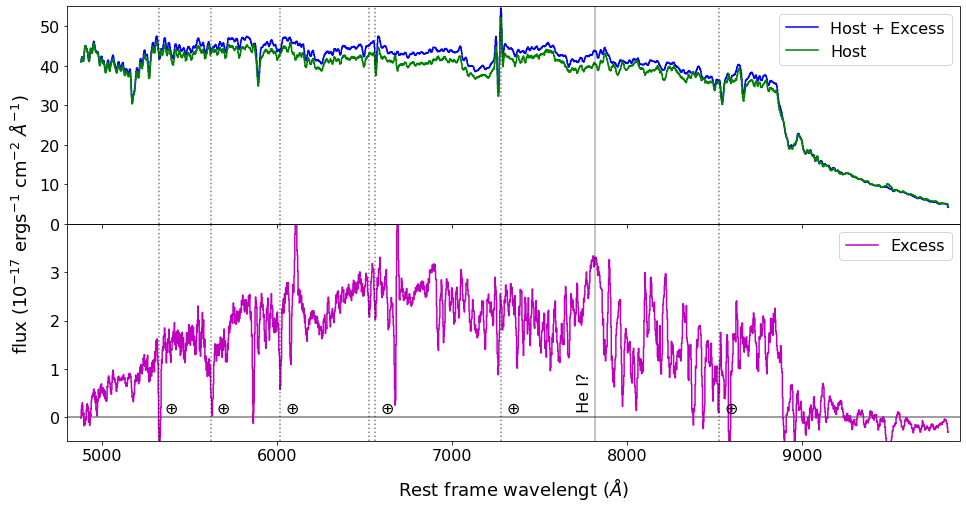}
    \caption{GTC spectrum of the transient in the SN 2019zbq host galaxy, taken on 4 December 2019. Vertical dotted lines give the locations of sky lines and tellurics. \textbf{Top:} Extracted spectra at the centre of the trace (host + excess) and at the side of the extended host galaxy trace (host). The sharp decline around $9000$~\AA\ is due to the low sensitivity of the detector at these wavelengths, and has been used to scale the spectra such that their difference goes to 0. \textbf{Bottom:} The difference of the two spectra in the top panel, showing the isolated spectrum of the excess. The host spectrum has been re-scaled to ensure that the red edge is the same for both spectra.}
    \label{2019zbq_spec}
\end{figure*}

\section{Late-time spectra of SN 2020qxz}
\begin{figure*}[h!]
    \centering
    \includegraphics[width=\textwidth]{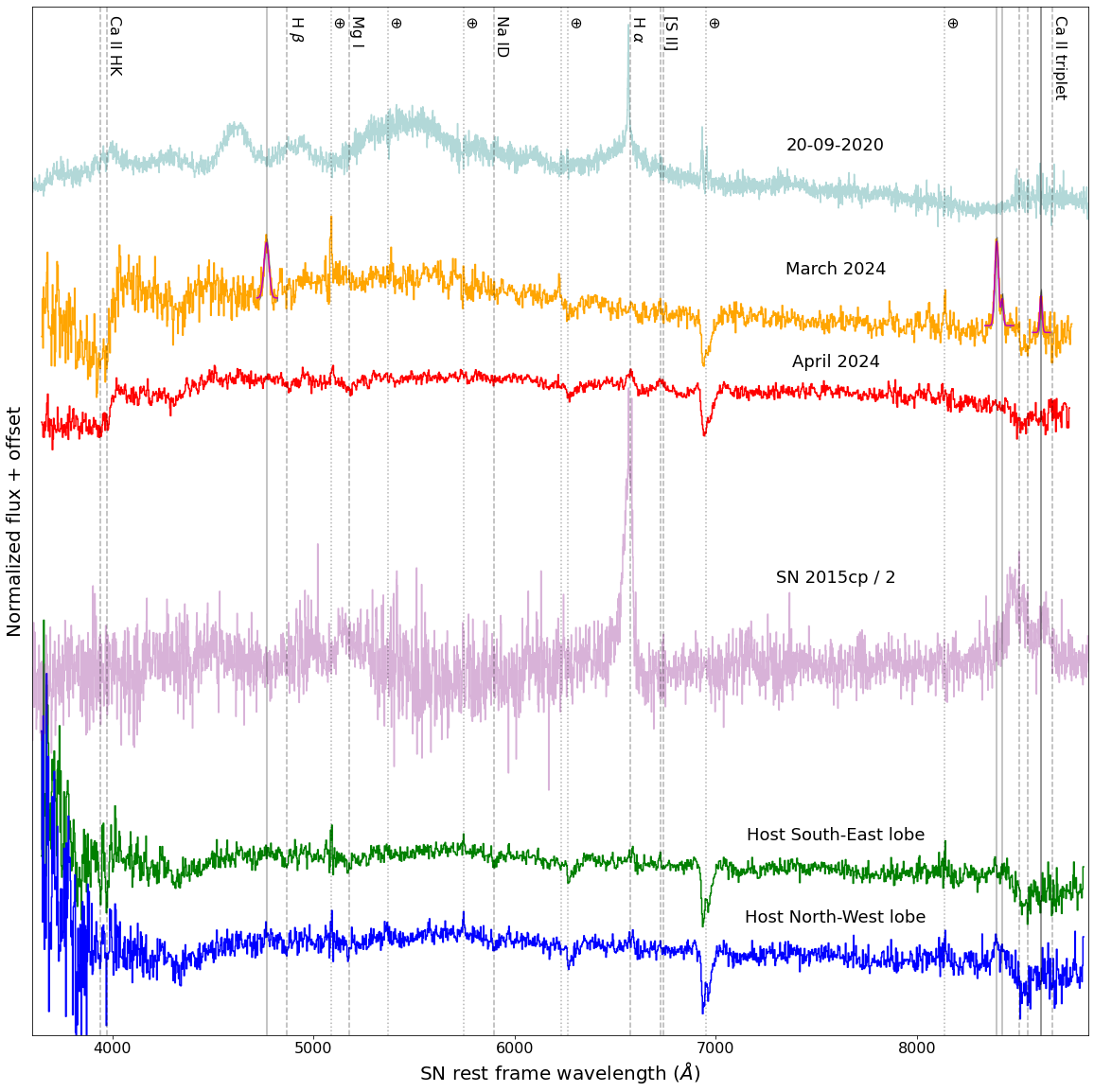}
    \caption{Same as Fig~\ref{2020qxz_spec_zoom} but showing the entire observed spectra. The transient emission lines are marked with grey vertical lines, and the best-fit Gaussians are overlaid on top of the combined March spectrum. The dashed vertical lines show the location of host galaxy lines (as well as \Hbeta\ to compare its location to the transient emission line) at the host redshift of $z=0.0975$, and the dotted vertical lines show the location of sky lines. We also show the classification spectrum of SN 2020qzx taken around peak magnitude in light blue and the late-time spectrum obtained for SN 2015cp in magenta, scaled down by a factor of 2 for readability.}
    \label{fig:2020qxz_spec}
\end{figure*}

\begin{figure*}
    \centering
    \includegraphics[width=\textwidth]{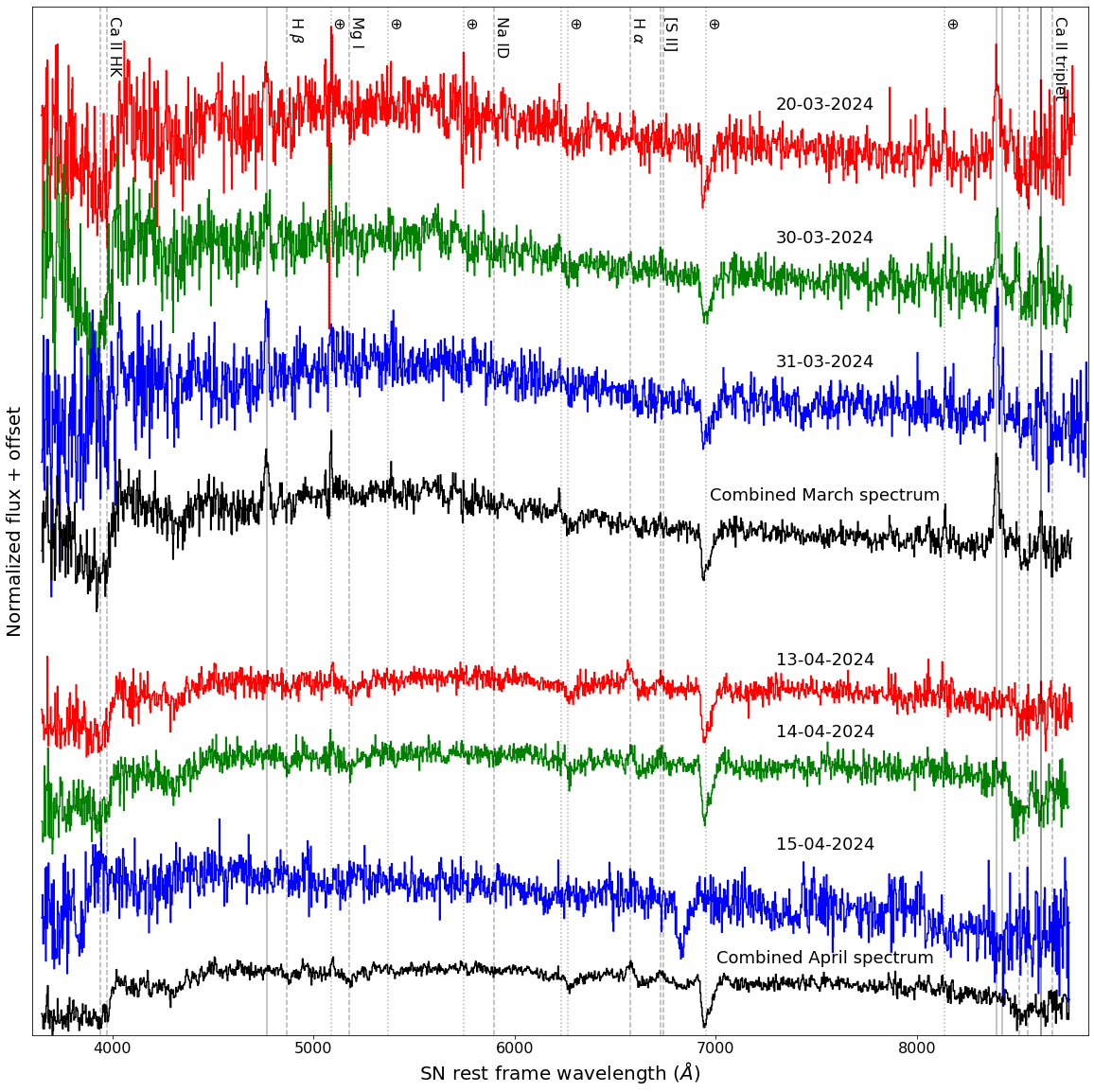}
    \caption{Individual late-time spectra of SN 2020qxz taken with the NOT. For both the March and April spectrum the three individual epochs are shown in red, green, and blue, and the stacked spectrum is shown in black. Each individual spectrum has an exposure time of 3 × 20 minutes, except for the spectrum taken on 15-04-2024 which is 2 × 20 minutes. The transient emission lines are marked with gray vertical lines, the dashed vertical lines show the location of host galaxy lines (as well as \Hbeta\ to compare its location to the transient emission line) at the host redshift of $z=0.0975$, and the dotted vertical lines show the location of sky lines.}
    \label{fig:2020qxz_spec_all_lines}
\end{figure*}

\begin{figure*}
    \centering
    \includegraphics[width=\textwidth]{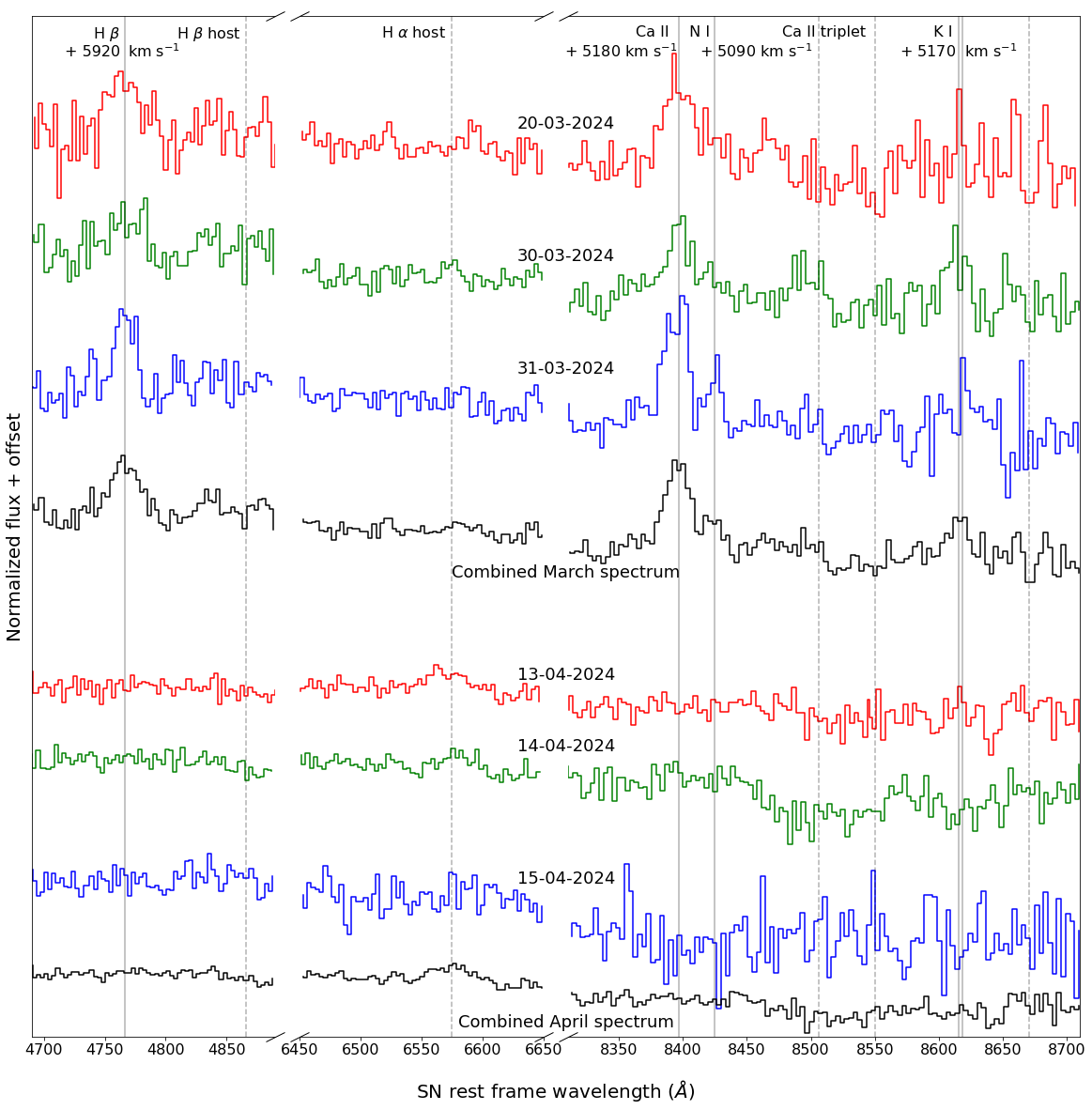}
    \caption{Same as Fig~\ref{fig:2020qxz_spec_all_lines} but zoomed in on the regions containing the transient emission lines and the \Halpha\ region.}
    \label{2020qxz_spec_zoom_all_lines}
\end{figure*}

\end{appendix}
\end{document}